\documentclass[submission,copyright,creativecommons]{eptcs}
\providecommand{\event}{EXPRESS/SOS 2023}

\ifpdf
  \usepackage{underscore}         
  \usepackage[T1]{fontenc}        
\else
  \usepackage{breakurl}           
\fi

\usepackage{amsmath,amsthm,amssymb}
\usepackage{mathpartir}
\usepackage{stmaryrd}
\usepackage{cmll}
\usepackage{proof}
\usepackage{xcolor}

\newcommand{\standard}{standard}

\newcommand{\defeq}{\triangleq}
\newcommand{\wt}[1]{\widetilde{#1}}
\newcommand{\tl}[1]{\tilde{#1}}
\newcommand{\emp}{\emptyset}

\newcommand{\Didtext}[1]{\textsc{\footnotesize #1}}
\newcommand{\Did}[1]{(\Didtext{#1})}

\newcommand{\bang}{\boldsymbol{*}}
\newcommand{\p}{$\pi$-}
\newcommand{\nil}{\mathbf{0}}
\newcommand{\selection}[2]{#1\triangleleft {#2}}

\newcommand{\branching}[3]{#1\triangleright\{{#2}_i:#3_i\}_{i\in I}}

\newcommand{\out}[1]{\oc\langle #1\rangle}
\newcommand{\inp}[1]{\wn({#1})}
\newcommand{\res}[1]{({\boldsymbol \nu} #1)}
\newcommand{\pp}{\mathrel{|}}

\newcommand{\picase}[3]
{\mathbf{case}~#1~\mathbf{of}~\{ \mathnormal{l}_i \_ ({#2}) \triangleright {#3} \}_{i\in I}}

\newcommand{\vv}[2]{\mathnormal{l}_{#1}\_ #2}

\newcommand{\unpack}[4]{{\mathbf{  open }\ {#1}\  \mathbf{ as }\ {(#2;#3)}\  \mathbf {in}\  {#4}}}

\newcommand{\var}[1]{{\m n}(#1)}

\newcommand{\variant}[2]{\langle \mathnormal{#1}\_ #2 \rangle_{i\in I}}
\newcommand{\linto}{\stackrel{1}{\rightarrow}}

\newcommand{\select}[2]{\oplus\{\mathnormal{#1}_i:#2_i\}_{i\in I}}
\newcommand{\branch}[2]{\&\{\mathnormal{#1}_i:#2_i\}_{i\in I}}

\newcommand{\nilT}{{{\m end}}}
\newcommand{\rec}[1]{\mu {#1}}
 
\newcommand{\dual}[1]{\overline{#1}}

\newcommand{\ch}{\sharp}
\newcommand{\ltp}[2]{\ell_{\m{#1}} [#2]}

\newcommand{\ssub}{{\ <:\ }}
\newcommand{\lsub}{{\ \leq\ }}

\newcommand{\polytype}[2]{\langle #1; #2\rangle}

\newcommand{\funtype}[1]{#1 \to \sigma}
\newcommand{\linfuntype}[1]{#1\linto \sigma}

\newcommand{\bool}{{\m{Bool}}}

\newcommand{\intg}{{\m{Int}}}

\newcommand{\unit}{{\mathbf{\star}}}
\newcommand{\unitT}{{\m{Unit}}}

\newcommand{\m}[1]{\tt{#1}}

\newcommand{\seqof}[1]{\wt{#1}}
\newcommand{\stype}{\stypeT}
\newcommand{\stypeT}{t}
\newcommand{\stypeS}{s}
\newcommand{\update}[3]{#1\{#2\mapsto#3\}}

\newcommand{\dom}{{\m{dom}}}

\newcommand{\sem}[1]{{\llbracket #1\rrbracket}}
\newcommand{\encf}[1]{\llbracket #1 \rrbracket_{\mathnormal{f}}}
\newcommand{\f}[1]{{\mathnormal f}_{#1}}
\newcommand{\encT}[1]{\llbracket #1\rrbracket}
\newcommand{\enc}[2]{\llbracket #1 \rrbracket_{\update{f}{x,y}{#2}}}
\newcommand{\encEMP}[2]{\llbracket #1 \rrbracket_{\{x,y \mapsto {#2}\}}}
\newcommand{\encx}[2]{\llbracket #1 \rrbracket_{\update{f}{x}{#2}}}
\newcommand{\encxEMP}[2]{\llbracket #1 \rrbracket_{\{x\mapsto {#2}\}}}

\newcommand{\encyEMP}[2]{\llbracket #1 \rrbracket_{\{y\mapsto {#2}\}}}

\newtheorem{definition}{Definition}[section]
\newtheorem{example}[definition]{Example}

\newtheorem{theorem}[definition]{Theorem}
\newtheorem{corollary}[definition]{Corollary}
\newtheorem{proposition}[definition]{Proposition}

\definecolor{blue-violet}{rgb}{0.54, 0.17, 0.89}

\def \bila {\begin{color}{blue-violet}} 
\def \eila {\end{color}}

\def \bilablack {\begin{color}{black}} 
\def \eilablack {\end{color}}

\def \bluc {\begin{color}{magenta}}
\def \eluc {\end{color}}

\def \bcomila {\begin{color}{blue-violet} {\bf Ilaria says: } } 
\def \ecomila {\end{color}}

\def \bds {\begin{color}{teal}} 
\def \eds {\end{color}}

\title{EXPRESSing Session Types}

\author{Ilaria Castellani
\institute{INRIA, Universit\'e C\^ote d'Azur} 
\\\\
Luca Padovani
\institute{University of Camerino}
\and
Ornela Dardha
\institute{University of Glasgow}
\\\\
Davide Sangiorgi
\institute{University of Bologna, INRIA}
}

\def\titlerunning{EXPRESSing Session Types}
\def\authorrunning{I. Castellani, O. Dardha, L. Padovani \& D. Sangiorgi}

\begin{document}

\maketitle
\begin{abstract}
  To celebrate the 30th edition of EXPRESS and the 20th edition of SOS we
  overview how session types can be expressed in a type theory for the standard
  \p calculus by means of a suitable encoding.
  The encoding allows one to reuse results about the \p calculus in the context
  of session-based communications, thus deepening the understanding of sessions
  and reducing redundancies in their theoretical foundations.
  Perhaps surprisingly, the encoding has practical implications as well, by
  enabling refined forms of deadlock analysis as well as allowing session type
  inference by means of a conventional type inference algorithm.
\end{abstract}

\section{Origins  of EXPRESS: some personal memories}
\label{s:memories}
             

This year marks an important milestone in the history of the
EXPRESS/SOS workshop series. Before joining their destinies in 2012,
the two workshops EXPRESS and SOS had been running on their own since
1994 and 2004, respectively.  Hence, the EXPRESS/SOS'23 workshop in
Antwerp will constitute the 30th edition of EXPRESS and the 20th
edition of SOS.

Two of us (Ilaria Castellani and Davide Sangiorgi) were personally
involved in the very first edition of EXPRESS in 1998, and indeed,
they may be said to have carried the workshop to the baptismal font,
together with Robert de Simone and Catuscia Palamidessi. Let us recall
some facts and personal memories.  The EXPRESS workshops were
originally held as meetings of the European project EXPRESS, a Network
of Excellence within the Human Capital and Mobility programme,
dedicated to expressiveness issues in Concurrency Theory. This NoE,
which lasted from January 1994 till December 1997, gathered
researchers from several European countries and was particularly
fruitful in supporting young researchers' mobility across different
sites.  The first three workshops of the NoE were held in Amsterdam
(1994), Tarquinia (1995), and Dagstuhl (1996).  The fourth and final
workshop was held in Santa Margherita Ligure (1997). It was co-chaired
by Catuscia Palamidessi and Joachim Parrow, and stood out as a
distinctive event, open to external participants and organised as a
conference with a call for papers.
A few months after this workshop, in the first half of 1998, the
co-chairs of the forthcoming CONCUR'98 conference in Nice, Robert de
Simone and Davide Sangiorgi, were wondering about endowing CONCUR with
a satellite event (such events were still unusual at the time) in
order to enhance its attractiveness.  Moreover, Davide was sharing
offices with Ilaria, who had been the NoE responsible for the site of
Sophia Antipolis and was also part of the organising committee of
CONCUR'98. It was so, during informal discussions, that the idea of
launching EXPRESS as a stand-alone event affiliated with CONCUR was
conceived, in order to preserve the heritage of the NoE and give it a
continuation.
Thus the first edition of EXPRESS, jointly chaired by Catuscia and
Ilaria, took place in Nice in 1998, as the first and unique satellite
event of CONCUR. However, EXPRESS did not remain a lonely satellite
for too long, as other workshops were to join the orbit of CONCUR in
the following years (INFINITY, YR-CONCUR, SecCo, TRENDS, \ldots),
including SOS in 2004.  The workshop EXPRESS'98 turned out to be
successful and very well attended.  Since then, EXPRESS has been
treading its path as a regular satellite workshop of CONCUR, with a
new pair of co-chairs every year, each co-chair serving two editions
in a row. The workshop, which is traditionally held on the Monday
preceding CONCUR, has always attracted good quality submissions and
has maintained a faithful audience over the years.

Coincidentally, this double anniversary of EXPRESS/SOS falls in the 30th
anniversary of Kohei Honda's first paper on session types~\cite{H93}. For this
reason, we propose an overview of a particular expressiveness issue, namely the
addition of session types to process calculi for mobility such as the
$\pi$-calculus.

\section{Session types and their expressiveness: introduction}

Expressiveness is a key topic in the design and implementation of
programming languages and models.  The issue is particularly relevant
in the case of formalisms for parallel and distributed systems, due
to the breadth and variety of constructs that have been proposed.

Most importantly, the study of expressiveness has practical
applications.  If the behaviours that can be programmed by means of a
certain formalism $L_1$ can also be programmed using another formalism
$L_2$, then methods and concepts developed for the latter language
(e.g., reasoning and implementation techniques) may be transferred
onto the former one that, in turn, may be more convenient to use from
a programming viewpoint. An important instance is the case
when $L_2$ is, syntactically, a subset of $L_1$. Indeed the quest for
a ``minimal'' formalism is central in the work on expressiveness.

This paper is an overview of a particular expressiveness issue, namely the
addition of session types onto calculi for mobility such as the $\pi$-calculus.
We will review the encoding of binary session types onto the \standard\
$\pi$-calculus \cite{DGS12,DGS17},
based on an observation of Kobayashi \cite{koba}. The key idea of the encoding is to represent a
sequence of communications within a session as a chain of
communications on linear channels (channels that are meant to be used
exactly once) through the use of explicit continuations, a technique
that resembles the modelling of communication patterns in the actor
model~\cite{Hewitt77}.  We discuss extensions of the encoding to
subtyping, polymorphism and higher-order communication as well as
multiparty session types. Finally, we review two applications of the
encoding to the problems of deadlock analysis and of session type
inference.

\emph{Session types}, initially proposed in \cite{H93,kubo1,kubo2},
describe \emph{sessions}, i.e., interaction protocols in distributed
systems.  While originally designed for process calculi,
they have later been integrated also in other paradigms,
including (multi-threaded) functional
programming~\cite{VGR06,pucellatov08,lindleymorris16,orchardyoshida17,
  KDLM21,KD21}, component-based systems~\cite{VVR06}, object-oriented
languages~\cite{DGDY06,DMYD06,CCDDG09}, languages for Web Services and
Contracts~\cite{CHY07,MY13}.
They have also been studied in logical-based type systems
\cite{CairesP10,W12,CairesPT12,DardhaG18,KokkeD21}.

Session types allow one to describe the sequences of input and output
operations that the participants of a session are supposed to follow,
explicitly indicating the types of messages being transmitted. This
structured \emph{sequentiality} of operations makes session types
suitable to model protocols.  Central (type and term) constructs in
session types are also branch and select, the former being the
offering of a set of alternatives and the latter being the selection
of one of the possible options at hand.

Session types were first introduced in a variant of the $\pi$-calculus
to describe binary interactions.  Subsequently, they have been
extended to {\em multiparty sessions}~\cite{HYC08}, where several
participants interact with each other. In the rest of this paper, we
will focus on \emph{binary session types}.

Session types guarantee privacy and communication safety within a
session.  Privacy means that session channels are known and used only
by the participants involved in the session.  Communication safety
means that interaction within a session will proceed without
mismatches of direction and of message type.  To achieve this, a
session channel is split into two endpoints, each of which is owned by
one of the participants. These endpoints are used according to dual
behaviours (and thus have dual types), namely one participant sends
what the other one is expecting to receive and vice versa.  Indeed,
\emph{duality} is a
key concept in the theory of session types.


To better understand session types and the notion of duality, let us
consider a simple example: the \emph{equality test}.  A \emph{server}
and a \emph{client} communicate over a session channel. The endpoints
$x$ and $y$ of the session channel are owned by the server and the
client, respectively and exclusively, and must have dual types. To
guarantee duality of types, static checks are performed by the type
system.

If the type of the server endpoint $x$ is
$$S \defeq \wn\intg.\wn\intg. \oc\bool.\nilT$$ --- meaning that the
process owning the channel endpoint $x$ receives ($\wn$) an integer
value followed by another integer value and then sends ($\oc$) back a
boolean value corresponding to the equality test of the integers
received --- then the type of the client endpoint $y$ should
be $$\overline S \defeq \oc\intg.\oc\intg. \wn\bool.\nilT$$ ---
meaning that the process owning the channel endpoint $y$ sends an
integer value followed by another integer value and then waits to
receive back a boolean value --- which is exactly the dual type.

There is a precise moment at which a session between two participants
is established. It is the \emph{connection} phase, when a fresh
(private) session channel is created and its endpoints are bound to
each communicating process.  The connection is also the moment when
duality, hence mutual compliance of two session types, is verified. In
order to establish a connection, 
primitives like {\tt accept/request} or $\res {xy}$, are added to the
syntax of terms \cite{kubo1,kubo2,fundamentals}.
%

%
When session types and session terms are added to the syntax of
\standard\ $\pi$-calculus types and terms, respectively, the syntax of
types (and, as a  consequence, of type environments)
usually needs to be split into two separate syntactic categories, one
for session types and the other for \standard\ $\pi$-calculus types
\cite{kubo1,kubo2,revisited,gayhole}.  Common typing features, like
subtyping, polymorphism, recursion have then to be added to both
syntactic categories.  Also the syntax of processes will contain both
\standard\ \p calculus process constructs and session process
constructs (for example, the constructs mentioned above to create
session channels).  These syntactic redundancies bring in redundancies
also in the theory, and can make the proofs of properties of the
language heavy.  Moreover, if a new type construct is added, the
corresponding properties must be checked both on \standard\ \p types
and on session types.  By ``\standard\ type systems'' we mean type
systems originally studied in depth 
for sequential languages such as the $\lambda$-calculus
and then transplanted onto the $\pi$-calculus as types for channel
names (rather than types for terms as in the $\lambda$-calculus); they
include, for instance, constructs for products, records, variants,
polymorphism, linearity, capabilities, and so on.

A further motivation for investigating the 
 expressiveness of the $\pi$-calculus with or without
 session types
 is the similarity between session constructs and \standard\
 $\pi$-calculus constructs.  Consider the type
 $S = \wn\intg.\wn\intg. \oc\bool.\nilT$. This type is assigned to a
 session channel endpoint and it describes a structured sequence of
 inputs and outputs by specifying the type of messages that the
 channel can transmit.  This way of proceeding reminds us of the
 \emph{linearised} channels~\cite{linearity}, which are channels used
 multiple times for communication but only in a sequential manner.
 Linearised types can, in turn, be encoded into linear types---i.e.,
 channel types used \emph{exactly once}~\cite{linearity}.  Similarly,
 there are analogies between the branch and select constructs of
 session types and the \emph{variant} types \cite{S98,pi} of
 \standard\ $\pi$-calculus types, as well as between the duality of
 session types, in which the behaviour of a session channel is split
 into two endpoints, and the \emph{capability types} of the \standard\
 $\pi$-calculus, that allow one to separate the input and output
 usages of channels.

In this paper we follow the encoding of binary session types into
linear \p types from \cite{DGS12,DGS17}, then discuss some extensions
and applications. The encoding was first suggested by Kobayashi
\cite{koba}, as a proof-of-concept without however formally studying
it. Later, Demangeon and Honda \cite{encoding} proposed an encoding of
session types into $\pi$-types with the aim of studying the subtyping
relation, and proving properties such as soundness of the encoding
with respect to typing and full abstraction.


\paragraph{Structure of the paper.} The rest of the paper is organised
as follows. In Section~\ref{s:back} we introduce the necessary
background about the session $\pi$-calculus and the linear
$\pi$-calculus. In Section~\ref{s:encoding} we recall the encoding
from the session $\pi$-calculus into the linear $\pi$-calculus, as
well as its correctness result.  In Section \ref{extensions} and
Section~\ref{sec:applications} we discuss respectively 
some extensions and some applications of the encoding.


\section{Background: $\pi$-calculus and session types}
\label{s:back}

In this section, we recall the syntax and semantics of our two
calculi of interest: the session $\pi$-calculus and the standard typed
$\pi$-calculus. We also introduce the notion of duality for
session types.

\begin{figure*}[t]
\centering
\begin{tabular}{@{}c@{}}
\begin{math}
	\displaystyle
	\begin{array}[t]{rlllll}
	T ::=	
		& S  	   		& \mbox{(session type)}\\
		& \ch T 		& \mbox{(channel type)}\\
		& \unitT 		& \mbox{(unit type)} \\
		& \ldots 		& \mbox{(other types)}\\
	\end{array}
	\begin{array}[t]{rlllll}
	S ::=  	
		& \nilT   			& \mbox{(termination)} \\
		& \oc T.S			& \mbox{(send)}      \\
		& \wn T.S		& \mbox{(receive)}	      \\
                & \select lS		& \mbox{(select)}\\
		& \branch lS		& \mbox{(branch)}

	\end{array}
\end{math}
\\\\ \hline \\
\begin{math}
	\displaystyle
	\begin{array}[t]{rllllll}
	P,Q ::= 	
		& x\out v.P				& \mbox{(output)} 				&
		& \nil	     				& \mbox{(inaction)}				\\ 
		& x\inp y.P				& \mbox{(input)}     			& 
		& P \pp Q  	    		  	& \mbox{(composition)} 			\\
		& \selection x {l_j}.P 		& \mbox{(selection)}  			&
		& \res {xy}P				& \mbox{(session restriction)}	\\
		& \branching xlP			& \mbox{(branching)}			& 
		& \res x P 				& \mbox{(channel restriction)}	\\
		\\
	v ::=
		& x 						& \mbox{(name)}					& 
		& \unit					& \mbox{(unit value)}
  \end{array}
\end{math}
\\\\ \hline \\
\begin{math}
	\displaystyle
    \begin{array}{rll}
    	\Did{R- StndCom}	&
    	& x\out {v}.P\pp x\inp {z}.Q \to P\pp Q[v/z]
      \\[1mm]    
	\Did{R-Com} && 
		\res {xy}(x\out v.P\pp y\inp z.Q) 
		\to 
		\res {xy}(P\pp Q[v/z])
      \\[1mm]      
	\Did{R-Case} &&
		\res {xy}(\selection x{l_j}.P \pp \branching ylP) 
		\to 
		\res {xy}(P\pp P_j)
	\quad  j\in I 
       \\[1mm]      
	\Did{R-StndRes}&&	{P\to Q} \Longrightarrow
							{\res {x} P\to \res {x} Q}
      \\[1mm]
      	\Did{R-Res}&	&{P\to Q} \Longrightarrow
							{\res {xy} P\to \res {xy} Q}
      \\[1mm]         
	\Did{R-Par}&&		{P\to Q}\Longrightarrow
							{P\pp R\to Q\pp R}
      \\[1mm]      
	\Did{R-Struct}&& 	{P\equiv P',\ P\to Q,\ Q'\equiv Q}
								\Longrightarrow
								{P'\to
                         Q'}
      \\
      \\
    \end{array}
\end{math}
 \end{tabular}
\caption{Syntax and
  reduction semantics of the session  \p calculus }
\label{f:syn}
\end{figure*}

\paragraph{Session types and terms.}
The syntax for session types and session \p calculus terms is reported
in Figure~\ref{f:syn}, together with the rules for the reduction
semantics, in which  
$\equiv$ is the usual \emph{structural congruence} relation, 
 allowing one to
rearrange parallel compositions and the scope of restrictions and
 to remove useless restrictions.
 We refer to, e.g., \cite{fundamentals,gayhole} for the rules for
 typing. 
Session types range over $S$ and types range over
$T$; the latter include session types, standard channel types denoted
by $\ch T$,
data types, such as $\unitT$ and any other type construct needed for mainstream
programming.

Session types are: $\nilT$, the type of a terminated channel; $ \wn
T.S$ and $ \oc T.S$ (used in the equality test example given in the
introduction) indicating, respectively, the {receive} and
{send} of a value of type $T$,  with
continuation type $S$. {Branch} and {select} are sets of labelled session types,
whose labels have indices ranging over a non-empty set
$I$.
Branch $\branch lS$ indicates an
{external choice}, namely what is offered, and it is a generalisation of the
input type in which the continuation $S_i$ \emph{depends} on the received label
$l_i$. Dually, select $\select lS$ indicates an {internal choice}, where only
one of the available labels $l_i$'s 
will be chosen, and it is a generalisation of the
output type.

Session processes range over $P,Q$. The
output process $x\oc \langle v\rangle .P$ sends a value $v$ on channel
endpoint $x$ and continues as $P$; the input process $x\wn(y).P$
receives on $x$ a value to substitute for the placeholder
$y$ in the continuation $P$. The selection process
$\selection x {l_j}.P$
selects label $l_j$ on channel $x$ and proceeds as
$P$. The branching process $\branching xlP$
offers a range of labelled alternative processes on channel
$x$. The session restriction construct $\res{xy}P$ creates a
session channel, more precisely its two endpoints $x$ and $y$, and
binds them in $P$. As usual, the term $\nil$ denotes a terminated
process and $P \pp Q$ the parallel composition of $P$ and $Q$.

\begin{figure}
	\centering
\begin{tabular}{@{}c@{}}
\begin{math}
	\displaystyle
	\begin{array}{rcllllllll}
		\stype ::=
			&\ltp{o}{\seqof\stype}		& \mbox{(linear output)}	&
			& \ch [\seqof\stype]			& \mbox{(connection)}		\\ 
			&\ltp{i}{\seqof\stype}		& \mbox{(linear input)}		&
			& \variant {l_i}{\stype_i} 		& \mbox{(variant type)}		\\
			& \ltp{\ch}{\seqof\stype}  	& \mbox{(linear connection)}&
			& \unitT 				  	& \mbox{(unit type)}		\\
			&\emptyset [] 				& \mbox{({no capability})}	&
			&\ldots 					& \mbox{(other types)}		\\
	\end{array}
\end{math}
\\\\ \hline\\
\begin{math}
	\displaystyle
	\begin{array}[t]{rlllll}
	P,Q ::= 	
			& x\out {\seqof v}.P		& \mbox{(output)} &
			& \nil				& \mbox{(inaction)} \\
			& x\inp {\seqof y}.P		& \mbox{(input)}     &
			& P \pp Q  	    		& \mbox{(composition)} \\
			& \res {x}P 		  	& \mbox{(restriction)} &
			& \picase{v}{x_i}{P_i}	& \mbox{(case)}\\\\
	v ::=	& x 						& \mbox{(name)} &
			& \unit				& \mbox{(unit value)}\\
			& \vv {}v				& \mbox{(variant value)}
  \end{array}
\end{math}
\\\\ \hline\\
\begin{math}
\begin{array}{rll} 
	\Did{R\p Com}&	& x\out {\tl v}.P\pp x\inp {\tl z}.Q \to P\pp Q[\tl v/\tl z]
      \\[1mm]    
	\Did{R\p Case}&	& \picase{\vv{j}v}{x_i}{P_i} \to P_j[v/x_j]\quad	j\in I
	\\[1mm]   
	\Did{R\p Res}& 	&{P\to Q} \Longrightarrow
							{\res {x} P\to \res {x} Q}
      \\[1mm]      
	\Did{R\p Par}& 	&{P\to Q}\Longrightarrow
							{P\pp R\to Q\pp R}
      \\[1mm]      
	\Did{R\p Struct}& &{P\equiv P',\ P\to Q,\ Q'\equiv Q}
								\Longrightarrow
								{P'\to
                            Q'}
  \\
  \\
\end{array}
\end{math}
 \end{tabular}
\caption{Syntax and reduction rules  of the standard typed $\pi$-calculus}
\label{f:pisyn}
\end{figure}

\paragraph{Duality}
Session type duality is a key ingredient in session types theory as it is
necessary for communication safety. Two processes willing to communicate, e.g.,
the client and the server in the equality test, 
must first agree on a session protocol. Intuitively, client and server 
should perform dual operations: when one process sends, the other receives, when one offers, the other chooses.  Hence, the dual of an input  must be an output, the dual of branch must be a select, and vice versa. Formally, duality on session types is defined as the following function:
\begin{displaymath}
  \begin{array}{rl}
	\dual{\nilT}				&\defeq\ \nilT\\
	\dual{\oc T.S}				&\defeq\ \wn T.\dual{S}\\
	\dual{\wn T.S}				&\defeq\ \oc T.\dual{S}\\
	\dual{\select lS}			&\defeq \ \branch l{\dual S} \\
	\dual{\branch lS}			&\defeq \ \select l{\dual S} \\
  \end{array}
\end{displaymath}

The static checks performed by the typing rules make sure that the peer
endpoints of the same session channel have dual types. In particular, this is
checked in the restriction rule $\Did{T-Res}$ below:
$$
\inferrule[{\Did{T-Res}}]
	{
	    \Gamma, x:T, y:\overline{T} \vdash P
	}
	{ 
	     \Gamma \vdash \res{xy} P
	}
$$

\paragraph{Standard $\pi$-calculus.}
The syntax and reduction semantics for the standard $\pi$-calculus are shown in
Figure~\ref{f:pisyn}. We use  $t$ to range over standard \p types, to
distinguish them from types $T$ and session types $S$, given in the previous
paragraph. We also use the notation $\seqof{\,\cdot\,}$ to indicate (finite)
sequences of elements. Standard \p types specify the \emph{capabilities} of
channels. The type $\emptyset []$ is assigned to a channel without any
capability, which cannot be used for any input/output action. Standard types
${\ltp{i}{\seqof\stype}}$ and ${\ltp{o}{\seqof\stype}}$ are assigned to channels
used \emph{exactly once} to receive and to send a sequence of values of type
$\seqof\stype$, respectively. The variant type $\variant {l_i}{\stype_i}$ is a
labelled form of disjoint union of types $\stype_i$ whose indices range over a
set $I$.

Linear types and variant types are essential in the encoding of
session types. The addition of variant types, as of any
structured type, implies the addition of a constructor in the grammar
for values, to produce variant values of the form $l\_v$, and of a
destructor in the grammar for processes, to consume variant
values. Such a destructor is represented by the term
$\picase{v}{x_i}{P_i}$, offering different behaviours depending on
which variant value $l\_v$ is received and binding $v$ to the
corresponding $x_i$. In the operational semantics, the reduction rule
in which a variant value is consumed \Did{R\p Case} 
 is sometimes called \emph{case
  normalisation}.
Unlike the session \p calculus, the standard \p calculus has just one
restriction operator that acts on single names, as in $\res x P$.

\section{Encoding sessions}
\label{s:encoding}

In this section we present the encoding of session \p calculus types
and terms into linear \p calculus types and terms, together with the main
 technical results, following 
 Dardha et al.~\cite{DGS12,DGS17}.
\begin{figure}
	\centering
	\begin{math}
		\displaystyle
		\begin{array}{rlllll}
			\encT{\nilT}		&\defeq&   \emp []									&\Did{E-End}\\
			\encT{\oc T.S}		&\defeq&   \ltp{o}{\encT{T},\encT{\dual S}}				&\Did{E-Out}\\
			\encT{\wn T.S}		&\defeq&   \ltp{i}{\encT{T},\encT{S}}					&\Did{E-Inp}\\
			\encT{ \select lS}	&\defeq&   \ltp{o}{\variant {l_i}{\encT{\dual{S_i}}}}			&\Did{E-Select}\\
			\encT{\branch lS}	&\defeq&   \ltp{i}{\variant {l_i}{\encT{S_i}}}				&\Did{E-Branch}\\
			\\\hline\\
			\encf{x}			&\defeq&  \f x			&\Did{E-Name}\\
			\encf{\star}		&\defeq&   \star			&\Did{E-Star}\\
			\\\hline\\
			\encf{\nil}			&\defeq&   \nil									&\Did{E-Inaction}\\
			\encf{x\out v.P}	&\defeq&   \res c \f x\out {\encf v,c}.\encx Pc			&\Did{E-Output}\\
			\encf{x\inp y.P}	&\defeq&   \f x\inp{y,c}.\encx Pc						&\Did{E-Input}\\[1mm]
			\encf{\selection x {l_j}.P}		&\defeq&    \res c \f x\out {\vv{j}c}.\encx Pc			&\Did{E-Selection}\\
			\encf{\branching xlP}	&\defeq&   \f x \inp y.\ \picase {y}{c}{\encx {P_i}c}		&\Did{E-Branching}\\
			\encf{P\pp Q}		&\defeq&   \encf{P} \pp \encf{Q}					&\Did{E-Composition}\\
			\encf{\res{xy}P}	&\defeq&   \res {c} \enc Pc							&\Did{E-Restriction}\\
			\encf{\res{x}P}		&\defeq&
                                                         \res{x}
                                                         \encf{P}
                                                                                                                                        &\Did{E-New}\\
                  \\
		\end{array}
	\end{math}
	\caption{Encoding of types, values and processes.}
	\label{f:encboth}\label{fig:enctypes}\label{fig:encterms}
\end{figure}

\paragraph{Type encoding.}
The encoding of session types into linear \p types is shown at the top of
Figure~\ref{f:encboth}.  Types produced by grammar $T$ are encoded in a homomorphic way, e.g.,
$\encT{\ch T} \defeq \ch \encT T$.
The encoding of $\nilT$ is a channel with no capabilities
$\emptyset []$ that cannot be used further. Type $\wn T.S$
is encoded as the linear input channel type carrying a pair of values
whose types are the encodings of $T$ and of $S$. The encoding of
$\oc T.S$ is similar except that the type of the second component of
the pair is the encoding of $\dual S$, since it describes the type of
a channel as it will be used by the receiver process. The branch and
the select types are encoded as linear input and linear output
channels carrying variant types having labels $l_i$ and types
that are respectively the encoding of $S_i$ and the encoding of
$\dual{S_i}$ for all $i \in I$. Again, the reason for using duality of
the continuation in the encoding of the select type
is the same as for the output type, as select is a generalisation of output type.

\paragraph{Process encoding.}
The encoding of session processes  into standard \p processes is shown at the
bottom of Figure~\ref{f:encboth}. The encoding of a process $P$ is parametrised
by a function $f$ from channel names to channel names.
We say that $f$ is a \emph{renaming function for $P$} if, for all the names $x$
that occur free in $P$, either $f(x) = x$ or $f(x)$ is a fresh name not
occurring in $\var P$, where $\var P$ is the set of all names of $P$, both free
and bound. Also, $f$ is the identity function on all bound names of $P$.
Hereafter we write $\dom(f)$ for the domain of $f$ and $\f x$ as an abbreviation
for $f(x)$. During the encoding of a session process, its renaming function $f$
is progressively updated. For example, we write $\update{f}{x}{c}$ or
$\update{f}{x,y}{c}$ for the update of $f$ such that the names $x$ and $y$ are
associated to $c$.
The notion of a renaming function is extended also to values as expected.
In the uses of the definition of the renaming function $f$ for $P$ (respectively
$v$), process $P$ (respectively value $v$) will be typed in a typing context,
say $\Gamma$. It is  implicitly assumed that the fresh names used by $f$ (that
is, the names $y$ such that $y=f(x)$, for some $x\neq y$) are also fresh for
$\Gamma$.

The motivation for parametrising the encoding of processes and values with a
renaming function stems from the key idea of encoding a structured communication
over a session channel as a chain of one-shot communications over \emph{linear}
channels. Whenever we transmit some payload on a linear channel, the payload is
paired with a fresh continuation channel on which the rest of the communication
takes place. Such continuation, being fresh, is different from the original
channel. Thus, the renaming function allows us to keep track of this fresh name
after each communication.


We now provide some more details on the encoding of terms.
Values are encoded as expected, so that a channel name $x$ is encoded to $\f{x}$
and the $\unit$ unit value is encoded to itself. This encoding is trivially extended to
every ground value added to the language.
In the encoding of the output process, a new channel name $c$ is created and is
sent together with the encoding of the payload $v$ along the channel $\f{x}$.
The encoding of the continuation process $P$ is parametrised by an updated $f$
where the name $x$ is associated to $c$. Similarly, the input process listens on
channel $\f{x}$ and receives a pair whose first element (the payload) replaces
the name $y$ and whose second element (the continuation channel $c$) replaces
$x$ in the continuation process by means of the updated renaming function
$\update{f}{x}{c}$.
As indicated in Section~\ref{s:back}, session restriction $\res {xy}P$
creates two fresh names and binds them in $P$ as the opposite endpoints of the
same session channel. This is not needed in the standard \p calculus. The
restriction construct $\res x P$ creates and binds a unique name $x$
in $P$;
this name identifies both endpoints of the communicating channel. The encoding
of a session restriction process $\res {xy}P$ is a standard channel restriction
process $\res {c}\enc Pc$ with the new name $c$ used to substitute both $x$ and $y$
in the encoding of $P$.
Selection $x \triangleleft {l}_j.P$ is encoded as the process that first creates
a new channel $c$ and then sends on $\f{x}$ a variant value $l_j\_c$, where
$l_j$ is the selected label and $c$ is the channel created to be used for the
continuation of the session. The encoding of branching receives on $\f{x}$ a
value, typically being a variant value, which is the guard of the {\bf case}
process. According to the transmitted label, one of the corresponding processes
$\encx {P_i}c$ for $i\in I$ will be chosen. Note that the name $c$ is bound in
any process $\sem{P_i}_{\update{f}{x}{c}}$.
The encoding of the other process constructs, namely inaction, standard channel
restriction, and parallel composition, acts as a homomorphism.

\newcommand{\server}{\mathit{server}}
\newcommand{\client}{\mathit{client}}

\begin{example}[Equality test]
	\label{ex:equality}
        We illustrate the encoding of session types and terms on the
        equality test from the introduction.  Thus we also make use of
        boolean and integer values, and simple operations on them,
        whose addition to the encoding is straightforward.

The encoding of the server's session type $S$ is
\[
	\encT S = \ltp{i}{\intg,\ltp{i}{\intg, \ltp o{\bool, \emp[ ]}}}
\]
while that of the client's session type $\overline S$ is
\[
	\encT{\overline S} = \ltp{o}{\intg,\ltp{i}{\intg, \ltp o{\bool, \emp[ ]}}}
\]

Note how the encoding of dual session types boils down to linear channel types
that have the \emph{same} payload and dual \emph{outermost} capabilities
${\ltp{i}{\cdot}}$ and ${\ltp{o}{\cdot}}$. This property holds in general and
can be exploited to express the (complex) notion of session type duality in
terms of the (simple) property of type equality, as we will see in
Section~\ref{sec:applications}.

The server
process, communicating on endpoint $x$ of type $S$,
is
$$\server \defeq x\inp {z_1}.x\inp {z_2}.x\out{z_1==z_2}.\nil$$ 
and 
 the client process, communicating on endpoint $y$ of
type $\overline S$,
is
 $$\client \defeq y\out 3
.y\out{5}.y\inp{eq}.\nil$$  Then we have
\[
\begin{array}{rcl}
  \encxEMP{\server}s & = & s\inp{z_1,c}. \llbracket {x\inp
    {z_2}.x\out{z_1==z_2}.\nil}\rrbracket _{
\{x\mapsto c\}}
\\[5pt]
& = &
 s\inp{z_1,c}.c\inp{z_2,c'}. \res {c''} c'\out{ z_1==z_2, c''}.\nil
\\

\end{array}
       \]
Similarly, 
\[
	\encyEMP{\client}s = \res c s\out{3,c}.\res {c'} c\out{5,c'}.c'\inp{eq,c''}.\nil
\]

The whole server-client system is thus encoded as follows, 
using  $\emptyset$ for 
      the identity function.
\[
	\sem{\res{xy} (\server \pp \client)}_\emptyset
	= \res{s} \encEMP{(\server \pp \client)}{s}
	= \res s \left(\encxEMP{\server}s \pp \encyEMP{\client}s \right)
      \]
(The update
      $\{ x,y \mapsto s\}$ reduces to $\{ x \mapsto s\}$ on the server
      and to $\{ y \mapsto s\}$ on the client because they do not
      contain occurrences of $y$ and $x$ respectively.)
%
\end{example}


\newcommand{\encynew}[2]{\llbracket #1 \rrbracket_{\update{f}{y}{#2}}}

\paragraph{Correctness of the encoding.}
The presented encoding can be considered as a semantics of session types and
session terms. The following theoretical results show that indeed we can derive the typing
judgements and the properties of the \p calculus with sessions via the encoding
and the corresponding properties of the linear \p calculus.

First, the correctness of an encoded typing judgement on the target terms
implies the correctness of the judgement on the source terms, and conversely.
Similar results hold for values. The encoding is  extended to session typing
contexts in the expected manner.

\begin{theorem} [Type correctness]\label{thm:corrtyp}
The following properties hold:
\begin{enumerate}
	\item If\/ $\Gamma \vdash P$, then $\encf{\Gamma} \vdash \encf{P}$  for some renaming function $f$ for $P$;
	\item If\/  $\encf{\Gamma} \vdash \encf{P}$ for some renaming function $f$ for
$P$, then $\Gamma \vdash P$.
\end{enumerate}
\end{theorem}

Theorem~\ref{thm:corrtyp}, and more precisely its proof
\cite{DGS17,Dardha16}, shows that the encoding can be actually used to
reconstruct the typing rules of session types.
That is, the typing rules for an operator {\tt op} of the session
$\pi$-calculus can be `read back' from the typing of the encoding of
{\tt op}.
%

Next we recall the operational correctness of the encoding. That is, the
property that the encoding allows one to faithfully reconstruct the behaviour of
a source term from that of the corresponding target term.  
We recall that $\to$  is the reduction relation of the two calculi.
We write $\hookrightarrow$ for
the extension of 
the structural congruence 
 $\equiv$ with a \emph{case normalisation} indicating the
decomposition of a variant value (Section~\ref{s:back}).

\begin{theorem}[Operational correspondence]\label{thm:oc} Let $P$ be a session
  process, $\Gamma$ a session typing context, and $f$ a renaming function for
  $P$ such that $\encf \Gamma \vdash \encf P$. Then the following statements
  hold.
  \begin{enumerate}
  \item
	 If\/ $P\to P'$, then $\encf P\to\hookrightarrow\encf {P'}$.
 \item 
	If\/ $\encf P\to Q$, then there is a session process $P'$ such that
	\begin{itemize}
	\item either $P\to P'$;
	\item or there are $x$ and $y$ such that $\res {xy}P\to P'$
	\end{itemize}
	and $Q\hookrightarrow \encf{P'}$.
  \end{enumerate}
\end{theorem}
Statement $1$ of the above theorem tells us that the reduction of an
encoded process mimics faithfully the reduction of the source process,
modulo structural congruence or case normalisation.  Statement $2$ of
the theorem tells us that if the encoding of a process $P$ reduces to
the encoding of a process $P'$ (via some intermediate process $Q$),
then the source process $P$ will reduce directly to $P'$ or it might
need a wrap-up under restriction. The reason for the latter is that in
the session \p calculus \cite{fundamentals}, reduction only occurs
under restriction and cannot occur
along free names. In particular, in the theorem, $f$ is a generic
renaming function; this function could map two free names, say $x$ and
$y$, onto the same name; in this case, an input at $x$ and an output
at $y$ in the source process could not produce a reduction, whereas
they might in the target process.
The two theorems
above allow us to derive, as
a straightforward corollary, the subject reduction property for the session
calculus.

\begin{corollary}[Session Subject Reduction]
\label{c:subjred}
If\/ $\Gamma\vdash P$ and $P\to Q$, then $\Gamma\vdash Q$.
\end{corollary}
 
Other properties of the session \p calculus can be similarly derived from
corresponding properties of the standard\ $\pi$-calculus. For instance, since
the encoding respects structural congruence (that is, $P\equiv P'$ if and only
if $\encf P\equiv \encf{P'}$), we can derive the invariance of typing under
structural congruence in the session \p calculus.

\begin{corollary}[Session Structural Congruence]
\label{c:typepressess}
If\/ $\Gamma\vdash P$ and $P\equiv P'$, then also $\Gamma\vdash P'$.
\end{corollary}


 

\newcommand{\stF}[1]{{#1}}
\newcommand{\gtF}[1]{{#1}}
\newcommand{\participant}[1]{\ensuremath{\gtF{\mathtt{#1}}}}
\newcommand{\role}{\ensuremath{\participant{p}}}
\newcommand{\q}{\ensuremath{\participant{q}}}
\newcommand{\pro}[2]{\ensuremath{\stF{\gtF{#1}\mathop{\upharpoonright}\gtF{#2}}}}

\section{Extensions}
\label{extensions}

In this section we discuss several extensions for the presented encoding,
which have been proposed in order to accommodate the additional
features of subtyping, polymorphism, recursion, higher-order
communication and multiparty interactions.

\paragraph{Subtyping.}
Subtyping is a relation between types based on a notion of
substitutability.  
If $T$ is a subtype of $T'$,
then any channel of type $T$ can be safely used in a context where a channel of
type $T'$ is expected.
In the standard $\pi$-calculus, subtyping originates from
\emph{capability types}~--- the possibility of distinguishing the
input and output usage of channels~\cite{sub,pi}. (This is analogous
to what happens in languages with references, where capabilities are
represented by the read and write usages.)  Precisely, the input
channel capability is co-variant, whereas the output channel
capability is contra-variant in the types of values transmitted (the
use of capabilities is actually necessary with linear types, as
reported in Figure~\ref{f:pisyn}). Subtyping can then be enhanced by
means of the variant types, which are co-variant both in depth and in
breadth.  In the case of session $\pi$-calculus, subtyping must be
dealt with also at the level of session types~\cite{gayhole}; for
instance, branch and select are both co-variant in depth, whereas they
are co-variant and contra-variant in breadth, respectively. This
duplication of effort can become heavy, particularly when types are
enriched with other constructs (a good example are recursive types).
The encoding of session types naturally accommodates subtyping, indeed
subtyping of the standard $\pi$-calculus can be used to derive
subtyping on session types.  Writing $\ssub$ and $\lsub$ for,
respectively, subtyping for session types and for standard \p types,
for instance we have:
\begin{theorem} [Encoding for Subtyping]
$T \ssub T'$ if and only if $\encT{T} \lsub \encT{T'}$.
\end{theorem}

\paragraph{Polymorphism and Higher-Order Communication.}
\emph{Polymorphism} is a common and useful type abstraction in
programming languages, as it allows operations that are generic by
using an expression with several types.  Parametric polymorphism has
been studied in the standard $\pi$-calculus~\cite{pi}, and in the \p
calculus with session types~\cite{CPPT13}; for bounded polymorphism in
session \p calculus see Gay \cite{boundpoly}.

The \emph{Higher-Order \p calculus} (HO$\pi$) models mobility of
processes that can be sent and received and thus can be run
locally~\cite{pi}.  Higher-order communication for the session \p
calculus \cite{ho} has the same benefits as for the \p calculus, in
particular, it models code mobility in a distributed scenario.

Extensions of the encoding to support polymorphism and HO$\pi$ have
been studied in \cite{DGS12,DGS17,Dardha16} and used to test its
robustness. The syntax of types and terms is extended to accommodate
the new constructs.
For polymorphism, session types and standard \p types are extended
with a {\em type variable} $X$ and with {\em polymorphic types}
$\polytype{X}{T}$ and $\polytype{X}{t}$, respectively.  For
higher-order communication, session types and standard \p types are
extended with the functional type $\funtype{T}$, assigned to a
functional term that can be used without any restriction, and with the
linear functional type $\linfuntype{T}$ that must be used exactly
once.  Correspondingly, the syntax of processes 
is extended to accommodate the \emph{unpacking} process
($\unpack{v}{X}{x}{P}$) to deal with polymorphism, and with
call-by-value $\lambda$-calculus primitives, namely \emph{abstraction}
($\lambda x:T. P$) and \emph{application} ($PQ$), to deal with
higher-order communication.

The encoding of the new type and 
process constructs is a homomorphism in all cases.
Consequently, the proof cases added to Theorems \ref{thm:corrtyp} and \ref{thm:oc} are trivial.

\paragraph{Recursion.}
The encoding was also extended to accommodate recursive types and replicated
processes by Dardha~\cite{D14}.  Here, the new added types are a recursive type
$\rec X.T$ and a type variable $X$, as well as the replicated process $\bang P$.
Recursive (session) types are required to be {\em guarded}, meaning that in
$\rec X.T$, variable $X$ may occur free in $T$ only under at least one of the
other type constructs.  The paper uses a new duality function, called
\emph{complement}, which is inspired by the work of Bernardi et
al.~\cite{BH13,BDGK14}.  Some new cases for the encoding of recursive session
types and 
processes are:
\begin{displaymath}
\begin{array}{llcl}
	\encT{X}				&\defeq\ X								\\
	\encT{\rec X.S}		&\defeq\ \rec X.{\encT{S}} \\
	\encf{*P}				&\defeq\ \ *\encf{P}		
\end{array}
\end{displaymath}
The extended encoding is proved to be sound and complete with respect to typing and reduction (aka operational correspondence). We refer the interested reader to \cite{D14,D14Ext}.

\paragraph{Multiparty Session Types.}
Multiparty Session Types (MPSTs)~\cite{HYC08,HYC16} accommodate communications
between more than two participants. Since their introduction, they have become a
major area of investigation within the session type community. Their meta-theory
is more complex than that of the binary case, and it is beyond the scope of this
paper to revise it in detail.
%
%

The core syntax of \emph{multiparty session types} is given by the following
grammar
\[
  	\begin{array}{r@{\hskip 2mm}c@{\hskip 2mm}l@{\hskip 3mm}l}
			\stF{S} &::=	& \nilT \pp X \pp
                                       \rec X.S & \mbox{(termination, type variable, recursive type)}
      		\\
          			&		& \role {\oplus_{i\in I}}{\oc{l}_i(U_i).S_i}	& \mbox{(select towards role $\role$)}
		\\
				&		& \role {\&_{i\in I}}{\wn{l}_i(U_i).S_i}	& \mbox{(branch from role $\role$)}\\
		\\
		\multicolumn{4}{c}{%
			\stF{B} \;::=\;
					\unitT \;|\; \ldots 			\quad\mbox{(base type)}
		\qquad%
			U \;::=\;
					\stF{B} \;|\; \stF{S}\ \mbox{(closed under $\mu$)}	\quad\mbox{(payload type)}
		}
	\end{array}
\]
where selection and branching types are annotated with \emph{roles}
identifying the participant of a multiparty session to which a message
is sent or from which a message is expected.
The message consists of a label $l_i$ and a payload of type $U_i$, whereas the
continuation $S_i$ indicates how the session endpoint is meant to be used
afterwards.


A multiparty session type describes the behaviour of a participant of a
multiparty session with respect to all the other participants it interacts with,
identified by their role in the session type. In order to obtain the behaviour
of a participant with respect to another \emph{particular} participant of the
multiparty session, say $\q$, the multiparty session type must be
\emph{projected} onto $\q$.
Hereafter, we write $\pro S\q$ for the \emph{partial projection of $\stF{S}$
onto $\q$},  referring to \cite{SDHY17,TECHREPORT} for its precise definition.
Projection yields a type defined by the following syntax, which resembles that
of binary session types:
\[
	\begin{array}{rcll}
		\stF{H} &::=& \nilT \pp X \pp \rec X.H & \mbox{(termination, type variable, recursive type)} \\
			&&{\oplus_{i\in I}}{\oc{l}_i(U_i).H_i}
                            & \mbox{(select)}\\
          	&&{\&_{i\in I}}{\wn{l}_i(U_i).H_i}	& \mbox{(branch)}
	\end{array}
\]

Projection is a key feature of MPSTs as it is needed in the technical
development of a sound type system. At the same time, it also provides a hook by
which multiparty sessions and multiparty session types can be encoded in the
standard \p calculus through the encoding of (binary) session types that we have
outlined in Section~\ref{s:back}.

Let us briefly comment on the encoding of MPST into linear types given
by Scalas et al.  \cite{SDHY17,TECHREPORT}.  This encoding is fully
fledged as it covers the whole MPST and it preserves the theory's
\emph{distributivity}.  Previous work by Caires and P{\'{e}}rez
\cite{CP16} presents an encoding of MPST into binary session types via
a \emph{medium} process, which acts as an orchestrator for the
encoding, thus losing distributivity.  In the encoding of Scalas et
al. no orchestrator is used, hence the encoding preserves its intended
choreographic nature as opposed to being orchestrated.

The encoding of a multiparty session type from Scalas et al. is formally defined as:
\[
	\encT{S} \defeq [\role : \encT{\pro S\role}]_{\role\in S}
\]
resulting in a \emph{record} of types with an entry \emph{for each role} {\role}
occurring in the multiparty session type $S$.
The encoding of a projected type, namely $\encT{$\pro S\role$}$, can then be
obtained by suitably adapting the function defined in Figure~\ref{f:encboth}.
The main cases are summarised below, and the encoding is a
homomorphism for the other
constructs in the syntactic category $H$ presented above.
\[
  \begin{array}{rcll}
	&\encT{{\oplus_{i\in I}}{\oc{l}_i(U_i).H_i}}	&\defeq \ltp{o}{\variant {l_i}{(\encT{U_i},\encT{H_i})}}
    \\[1mm]
	&\encT{{\&_{i\in I}}{\wn{l}_i(U_i).H_i}}	&
                                                           \defeq
                                                           \ltp{i}{\variant
                                                           {l_i}{(\encT{U_i},\encT{H_i})}}
  \end{array}
\]

The encoding of processes is quite complex and beyond the scope of
this paper.  The interested reader may refer to Scalas et
al.~\cite{SDHY17,SDHY17b} for the formal details and a Scala
implementation of multiparty sessions based on this encoding.  The
encoding of MPST into linear types satisfies several properties,
including duality and subtyping preservation, correctness of the
encoding with respect to typing, operational correspondence and
deadlock freedom preservation. These properties are given in Section 6
of \cite{SDHY17}.

\section{Applications}
\label{sec:applications}

The encoding from session types to linear channel types can be thought of as a
way of ``explaining'' a high-level type language in terms of a simpler,
lower-level type language. Protocols written in the lower-level type language
tend to be more cumbersome and less readable than the session types they encode.
For this reason, it is natural to think of the encoding as nothing more than a
theoretical study. Yet, as we are about to see in this section, the very same
encoding has also enabled (or at least inspired) further advancements in the
theory and practice of session types.

\subsection{A Type System for Deadlock Freedom}

A well-typed session \p-calculus process (and equivalently a well-typed standard
\p-calculus one) enjoys communication safety (no message with unexpected type is
ever exchanged) but not deadlock freedom. For example, both the session
\p-calculus process
\begin{equation}
    \label{eq:deadlock2}
    \res{x_1x_2}\res{y_1y_2}\left(
    x_1\inp{z}. y_1\out{z}.\nil \pp
    y_2\inp{w}.x_2\out{w}.\nil
    \right)
\end{equation}
and the standard \p-calculus process
\begin{equation}
    \label{eq:deadlock}
    \res{x}\res{y}\left(
    x\inp{}.y\out{}.\nil \pp
    y\inp{}.x\out{}.\nil
    \right)
\end{equation}
are well-typed in the respective typing disciplines, but the behaviours they
describe on the two sessions/channels are intermingled in such a way that no
communication can actually occur: the input from each session/channel must be
completed in order to perform the output on the other session/channel.
%
%
%

Several type systems that ensure deadlock freedom in addition to communication
safety have been studied for session and standard  typed \p calculi. In a
particular line of work, Kobayashi~\cite{K02,deadlock2} has studied a typing
discipline that associates \emph{priorities} to channel types so as to express,
at the type level, the relative order in which channels are used, thus enabling
the detection of circular dependencies, such as the one shown above. Later on,
Padovani~\cite{Padovani14} has specialised this technique for the linear
\p-calculus and, as an effect of the encoding illustrated in
Section~\ref{s:back}, for the session \p-calculus as well.
To illustrate the technique, in this section we consider a refinement
of the linear input/output types in Figure~\ref{f:pisyn}
as follows
\[
    \stype ::= \ltp{o}{\,\seqof\stype\,}^m \;\mid\; \ltp{i}{\,\seqof\stype\,}^n \;\mid\; \cdots
\]
where $m$ and $n$ are integers representing priorities: the smaller the number,
the higher the priority with which the channel must be used.
In the process~\eqref{eq:deadlock} above, we could assign the types
$\ltp{i}{}^m$ and $\ltp{o}{}^n$ to respectively $x$ and $y$ on the lhs of $\pp$
and the types $\ltp{o}{}^m$ and $\ltp{i}{}^n$ to respectively $x$ and $y$ on the
rhs of $\pp$.
Note that each channel is assigned 
two types having \emph{dual polarities} (each channel is used in
complementary ways on the two sides of $\pp$) and the \emph{same
  priority}. Then, the type system imposes constraints on priorities
to reflect the order in which channels are used: on the lhs of $\pp$
we have the constraint $m < n$ since the input on $x$ (with priority
$m$) blocks the output on $y$ (with priority $n$); on the rhs of $\pp$
the opposite happens, resulting in the constraint $n < m$. Obviously,
these two constraints are not simultaneously satisfiable, hence the
process as a whole is ruled out as ill typed.
%

In such simple form, this technique fails to deal with most recursive processes.
We illustrate the issue through the following server process that computes the
factorial of a natural number, in which we use a few standard extensions
(replication, conditional, numbers and their operations) to the calculus
introduced earlier.

\newcommand{\fact}{\mathit{fact}}
\newcommand{\If}[3]{\mathbf{if}~#1~\mathbf{then}~#2~\mathbf{else}~#3}

\begin{equation}
    \label{eq:fact}
    \bang\fact\inp{x,y}.
    \If{x=0}{
        y\out{1}
    }{
        \res{z}\left(
            \fact\out{x-1,z}
            \pp
            z\inp{k}.
            y\out{x \times k}
        \right)
    }
\end{equation}


The server accepts requests on a shared channel $\fact$. Each request carries a
natural number $x$ and a linear channel $y$ on which the factorial of $x$ is
sent as response. The modelling follows the standard recursive definition of the
factorial function. In particular, in the recursive case a fresh linear channel
$z$ is created from which the factorial $k$ of $x - 1$ is received. At that
point, the factorial $x \times k$ of $x$ can be sent on $y$.
Now assume, for the sake of illustration, that $m$ and $n$ are the priorities
associated with $y$ and $z$, respectively. Since $z$ is used in the same
position as $y$ in the recursive invocation of $\fact$, we expect that $z$ and
$y$ should have the same type hence the same priority $m = n$. This clashes with
the input on $z$ that blocks the output on $y$, requiring $n < m$.
The key observation we can make in order to come up with a more flexible
handling of priorities is that a replicated process like \eqref{eq:fact} above
cannot have any \emph{free} linear channel. In fact, the only free channel
$\fact$ is a shared one whereas $y$ is received by the process and $z$ is
created within the process. As a consequence, the absolute value of the
priorities $m$ and $n$ we associate with $y$ and $z$ does not matter (as long as
they satisfy the constraint $n < m$) and they can vary from one request to
another. In more technical terms, this corresponds to making $\fact$
\emph{polymorphic} in the priority of the channel $y$ received from it and
allowing a (priority-limited) form of \emph{polymorphic recursion} when we type
outputs such as $\fact\out{x-1,z}$.

It must be pointed out that a process such as \eqref{eq:fact} is in the scope of
Kobayashi's type systems~\cite{deadlock2}. The additional expressiveness
resulting from priority polymorphism enables the successful analysis of
recursive processes that interleave actions on different linear channels also in
cyclic network topologies. We do not showcase these more complex scenarios in
this brief survey, instead referring the interested reader to \cite{Padovani14}
for an exhaustive presentation of the technique and to~\cite{Padovani17} for a
proof-of-concept implementation.

As a final remark, it is interesting to note that this technique can
be retrofitted to a calculus with native sessions, but it was born in
the context of the standard \p-calculus, which features a more
primitive communication model.  The point is that, in the standard
\p-calculus, sequential communications are encoded in a
continuation-passing style, meaning that higher-order channels are the
norm rather than the exception. So, the quest for expressive type
systems ensuring (dead)lock freedom in the standard \p-calculus could
not ignore this feature, and this necessity has been a major source of
inspiration for the support of priority polymorphism.  In this
direction, Carbone et al. \cite{CDM14} study (dead)lock freedom for
session \p processes using the encoding from Section~\ref{s:encoding}
and the technique from \cite{deadlock2} and show that this combined
technique is more fine-grained than other ones adopted in session \p
calculi.  Dardha and P{\'{e}}rez \cite{DP22} present a full account of the deadlock freedom property in session \p calculi, and
compare deadlock freedom obtained by using the encoding and the work
from \cite{deadlock2} to linear logic approaches, which are used as a
yardstick for deadlock freedom.


\subsection{Session Type Inference}

A major concern regarding all type systems is their realisability and
applicability in real-world programming languages. In this respect, session type
systems pose at least three peculiar challenges: (1) the fact that session
endpoints must be treated as \emph{linear resources} that cannot be duplicated
or discarded; (2) the need to \emph{update} the session type associated with a
session endpoint each time the endpoint is used; (3) the need to express
\emph{session type duality} constraints in addition to the usual \emph{type
equality} constraints.
The first challenge can be easily dealt with only in those (few) languages that
provide native support for linear (or at least affine) types. Alternatively, it
is possible to devise mechanisms that detect linearity (or affinity) violations
at runtime with a modest overhead.
The second challenge can be elegantly addressed by adopting a \emph{functional}
API for sessions~\cite{GV10}, whereby each function/method using a session
endpoint returns (possibly along with other results) the same endpoint with its
type suitably updated.
The last challenge, which is the focus of this section, is a subtle one since
session type duality is a complex relation that involves the whole structure of
two session types. In fact, it has taken quite some time even just to
\emph{correctly define} duality in the presence of recursive session
types~\cite{BH13,GayThiemannVasconcelos20}.

Somewhat surprisingly, the encoding of session types into linear channel types
allows us to cope with this challenge in the most straightforward way, simply by
\emph{getting rid of it}.
In Example~\ref{ex:equality} we have shown two session types, one dual of the
other, whose respective encodings are \emph{equal} except for the outermost
capabilities. This property holds in general.

\begin{proposition}
    \label{prop:duality}
    Let $\overline{\,\cdot\,}$ be the partial involution on types such that
    $\overline{\emp[]} = \emp[]$ and $\overline{\ltp{i}{\,\seqof\stype\,}} =
    \ltp{o}{\,\seqof\stype\,}$ and $\overline{\ltp{o}{\,\seqof\stype\,}} =
    \ltp{i}{\,\seqof\stype\,}$. Then $\encT{\overline{S}} = \overline{\encT{S}}$
    for every $S$.
\end{proposition}

\newcommand{\emptyT}{\circ}
\newcommand{\fullT}{\bullet}

In fact, it is possible to devise a slightly different representation of
capabilities so that (session) type duality can be expressed solely in terms of
type equality. To this aim, let $\emptyT$ and $\fullT$ be any two types which we
use to represent the absence and presence of a given capability, respectively.
We do not need any particular property of $\emptyT$ and $\fullT$ except the fact
that they must be different. In fact, they need not even be inhabited. Now, we
can devise a slightly different syntax for linear channel types, as follows:
\[
    \stype ::= \ltp{\kappa,\kappa}{\,\seqof\stype\,} \mid \cdots
    \qquad
    \kappa ::= \emptyT \mid \fullT
\]

The idea is that a linear channel type carries two separate input and output
capabilities (hereafter ranged over by $\kappa$ and $\iota$), each of which can
be either present or absent. For example, $\ltp{\emptyT,\emptyT}{}$ would be the
same as $\emp[]$, $\ltp{\fullT,\emptyT}{\,\seqof\stype\,}$ would be the same as
$\ltp{i}{\,\seqof\stype\,}$ and $\ltp{\emptyT,\fullT}{\,\seqof\stype\,}$ would
be the same as $\ltp{o}{\,\seqof\stype\,}$.
With this representation of linear channel types the dual of a type can be
defined simply as $\overline{\ltp{\kappa,\iota}{\,\seqof\stype\,}} =
\ltp{\iota,\kappa}{\,\seqof\stype\,}$, where the input/output capabilities are
swapped. Now, suppose that we wish to express a duality constraint $S =
\overline{T}$ stating that $S$ is the dual of $T$ and let
$\ltp{\kappa,\iota}{\,\seqof\stypeS\,} = \encT{S}$ and
$\ltp{\kappa',\iota'}{\,\seqof\stypeT\,} = \encT{T}$ be the encodings of $S$ and
$T$, respectively. Using Proposition~\ref{prop:duality} and the revised
representation of linear channel types we obtain
\[
    S = \overline{T}
    \iff
    \kappa = \iota' \wedge \iota = \kappa' \wedge \seqof\stypeS = \seqof\stypeT
\]
thereby turning a session type duality constraint into a conjunction of type
equality constraints.

This apparently marginal consequence of using encoded (as opposed to native)
session types makes it possible to rely on completely standard features of
conventional type systems to express and infer complex structural relations on
session types. In particular, it allows any Hindley-Milner type inference
algorithm to perform \emph{session type inference}.
\texttt{FuSe}~\cite{Padovani17} is a library implementation of session types for
OCaml that showcases this idea at work. The library supports higher-order
sessions, recursive session types and session subtyping by piggybacking on
OCaml's type system. Clearly, the inferred (encoded) session types are not as
readable as the native ones. This may pose problems in the presence of type
errors. To address this issue, the library is accompanied by an external tool
called \texttt{Rosetta} that decodes encoded session types and pretty prints
them as native ones using the inverse of the encoding function
$\encT\cdot$.\footnote{The source code of \texttt{FuSe} and \texttt{Rosetta} is
  publicly available at \url{https://github.com/boystrange/FuSe}.}
On similar
lines, Scalas and Yoshida \cite{ScalasY16} develop \texttt{lchannels}, a Scala
library for session types fully based on the encoding of session types into
linear types. As a result, the structure of a session type is checked statically
by analysing its encoding onto channel types in Scala, while linearity is
checked dynamically at run time as in \texttt{FuSe}, as Scala has no support for
linearity.

\bibliographystyle{eptcs}
\bibliography{myBib,myBib2,luca,ilaria}

\begin{thebibliography}{10}
\providecommand{\bibitemdeclare}[2]{}
\providecommand{\surnamestart}{}
\providecommand{\surnameend}{}
\providecommand{\urlprefix}{Available at }
\providecommand{\url}[1]{\texttt{#1}}
\providecommand{\href}[2]{\texttt{#2}}
\providecommand{\urlalt}[2]{\href{#1}{#2}}
\providecommand{\doi}[1]{doi:\urlalt{https://doi.org/#1}{#1}}
\providecommand{\eprint}[1]{arXiv:\urlalt{https://arxiv.org/abs/#1}{#1}}
\providecommand{\bibinfo}[2]{#2}

\bibitemdeclare{inproceedings}{BDGK14}
\bibitem{BDGK14}
\bibinfo{author}{Giovanni \surnamestart Bernardi\surnameend},
  \bibinfo{author}{Ornela \surnamestart Dardha\surnameend},
  \bibinfo{author}{Simon~J. \surnamestart Gay\surnameend} \&
  \bibinfo{author}{Dimitrios \surnamestart Kouzapas\surnameend}
  (\bibinfo{year}{2014}): \emph{\bibinfo{title}{On Duality Relations for
  Session Types}}.
\newblock In: {\slshape \bibinfo{booktitle}{TGC}}, {\slshape
  \bibinfo{series}{LNCS}} \bibinfo{volume}{8902},
  \bibinfo{publisher}{Springer}, pp. \bibinfo{pages}{51--66},
  \doi{10.1007/978-3-662-45917-1_4}.

\bibitemdeclare{inproceedings}{BH13}
\bibitem{BH13}
\bibinfo{author}{Giovanni \surnamestart Bernardi\surnameend} \&
  \bibinfo{author}{Matthew \surnamestart Hennessy\surnameend}
  (\bibinfo{year}{2014}): \emph{\bibinfo{title}{Using Higher-Order Contracts to
  Model Session Types (Extended Abstract)}}.
\newblock In: {\slshape \bibinfo{booktitle}{CONCUR}}, {\slshape
  \bibinfo{series}{LNCS}} \bibinfo{volume}{8704},
  \bibinfo{publisher}{Springer}, pp. \bibinfo{pages}{387--401},
  \doi{10.1007/978-3-662-44584-6_27}.

\bibitemdeclare{inproceedings}{CP16}
\bibitem{CP16}
\bibinfo{author}{Lu{\'{\i}}s \surnamestart Caires\surnameend} \&
  \bibinfo{author}{Jorge~A. \surnamestart P{\'{e}}rez\surnameend}
  (\bibinfo{year}{2016}): \emph{\bibinfo{title}{Multiparty Session Types Within
  a Canonical Binary Theory, and Beyond}}.
\newblock In \bibinfo{editor}{Elvira \surnamestart Albert\surnameend} \&
  \bibinfo{editor}{Ivan \surnamestart Lanese\surnameend}, editors: {\slshape
  \bibinfo{booktitle}{FORTE}}, {\slshape \bibinfo{series}{LNCS}}
  \bibinfo{volume}{9688}, \bibinfo{publisher}{Springer}, pp.
  \bibinfo{pages}{74--95}, \doi{10.1007/978-3-319-39570-8\_6}.

\bibitemdeclare{inproceedings}{CPPT13}
\bibitem{CPPT13}
\bibinfo{author}{Lu{\'{\i}}s \surnamestart Caires\surnameend},
  \bibinfo{author}{Jorge~A. \surnamestart P{\'{e}}rez\surnameend},
  \bibinfo{author}{Frank \surnamestart Pfenning\surnameend} \&
  \bibinfo{author}{Bernardo \surnamestart Toninho\surnameend}
  (\bibinfo{year}{2013}): \emph{\bibinfo{title}{Behavioral Polymorphism and
  Parametricity in Session-Based Communication}}.
\newblock In: {\slshape \bibinfo{booktitle}{ESOP}}, {\slshape
  \bibinfo{series}{LNCS}} \bibinfo{volume}{7792},
  \bibinfo{publisher}{Springer}, pp. \bibinfo{pages}{330--349},
  \doi{10.1007/978-3-642-37036-6_19}.

\bibitemdeclare{inproceedings}{CairesP10}
\bibitem{CairesP10}
\bibinfo{author}{Lu{\'{\i}}s \surnamestart Caires\surnameend} \&
  \bibinfo{author}{Frank \surnamestart Pfenning\surnameend}
  (\bibinfo{year}{2010}): \emph{\bibinfo{title}{Session Types as Intuitionistic
  Linear Propositions}}.
\newblock In: {\slshape \bibinfo{booktitle}{Proc. of {CONCUR} 2010}}, {\slshape
  \bibinfo{series}{Lecture Notes in Computer Science}} \bibinfo{volume}{6269},
  \bibinfo{publisher}{Springer}, pp. \bibinfo{pages}{222--236},
  \doi{10.1007/978-3-642-15375-4\_16}.

\bibitemdeclare{article}{CairesPT12}
\bibitem{CairesPT12}
\bibinfo{author}{Lu\'{\i}s \surnamestart Caires\surnameend},
  \bibinfo{author}{Frank \surnamestart Pfenning\surnameend} \&
  \bibinfo{author}{Bernardo \surnamestart Toninho\surnameend}
  (\bibinfo{year}{2014}): \emph{\bibinfo{title}{Linear Logic Propositions as
  Session Types}}.
\newblock {\slshape \bibinfo{journal}{MSCS}}, \doi{10.1017/S0960129514000218}.

\bibitemdeclare{article}{CCDDG09}
\bibitem{CCDDG09}
\bibinfo{author}{Sara \surnamestart Capecchi\surnameend},
  \bibinfo{author}{Mario \surnamestart Coppo\surnameend},
  \bibinfo{author}{Mariangiola \surnamestart Dezani-Ciancaglini\surnameend},
  \bibinfo{author}{Sophia \surnamestart Drossopoulou\surnameend} \&
  \bibinfo{author}{Elena \surnamestart Giachino\surnameend}
  (\bibinfo{year}{2009}): \emph{\bibinfo{title}{Amalgamating sessions and
  methods in object-oriented languages with generics}}.
\newblock {\slshape \bibinfo{journal}{Theor. Comput. Sci.}}
  \bibinfo{volume}{410}(\bibinfo{number}{2-3}), pp. \bibinfo{pages}{142--167},
  \doi{10.1016/j.tcs.2008.09.016}.

\bibitemdeclare{inproceedings}{CDM14}
\bibitem{CDM14}
\bibinfo{author}{Marco \surnamestart Carbone\surnameend},
  \bibinfo{author}{Ornela \surnamestart Dardha\surnameend} \&
  \bibinfo{author}{Fabrizio \surnamestart Montesi\surnameend}
  (\bibinfo{year}{2014}): \emph{\bibinfo{title}{Progress as Compositional
  Lock-Freedom}}.
\newblock In: {\slshape \bibinfo{booktitle}{COORDINATION}}, {\slshape
  \bibinfo{series}{LNCS}} \bibinfo{volume}{8459},
  \bibinfo{publisher}{Springer}, pp. \bibinfo{pages}{49--64},
  \doi{10.1007/978-3-662-43376-8_4}.

\bibitemdeclare{inproceedings}{CHY07}
\bibitem{CHY07}
\bibinfo{author}{Marco \surnamestart Carbone\surnameend},
  \bibinfo{author}{Kohei \surnamestart Honda\surnameend} \&
  \bibinfo{author}{Nobuko \surnamestart Yoshida\surnameend}
  (\bibinfo{year}{2007}): \emph{\bibinfo{title}{Structured
  Communication-Centred Programming for Web Services}}.
\newblock In: {\slshape \bibinfo{booktitle}{ESOP}}, {\slshape
  \bibinfo{series}{{LNCS}}} \bibinfo{volume}{4421},
  \bibinfo{publisher}{{Springer}}, pp. \bibinfo{pages}{2--17},
  \doi{10.1007/978-3-540-71316-6_2}.

\bibitemdeclare{inproceedings}{D14}
\bibitem{D14}
\bibinfo{author}{Ornela \surnamestart Dardha\surnameend}
  (\bibinfo{year}{2014}): \emph{\bibinfo{title}{Recursive Session Types
  Revisited}}.
\newblock In: {\slshape \bibinfo{booktitle}{BEAT}}, {\slshape
  \bibinfo{series}{{EPTCS}}} \bibinfo{volume}{162}, pp.
  \bibinfo{pages}{27--34}, \doi{10.4204/EPTCS.162.4}.

\bibitemdeclare{misc}{D14Ext}
\bibitem{D14Ext}
\bibinfo{author}{Ornela \surnamestart Dardha\surnameend}
  (\bibinfo{year}{2014}): \emph{\bibinfo{title}{Recursive Session Types
  Revisited}}.
\newblock \bibinfo{note}{\\
  \url{http://www.dcs.gla.ac.uk/~ornela/my_papers/D14-Extended.pdf}}.

\bibitemdeclare{book}{Dardha16}
\bibitem{Dardha16}
\bibinfo{author}{Ornela \surnamestart Dardha\surnameend}
  (\bibinfo{year}{2016}): \emph{\bibinfo{title}{Type Systems for Distributed
  Programs: Components and Sessions}}.
\newblock {\slshape \bibinfo{series}{Atlantis Studies in
  Computing}}~\bibinfo{volume}{7}, \bibinfo{publisher}{Springer / Atlantis
  Press}, \doi{10.2991/978-94-6239-204-5}.

\bibitemdeclare{inproceedings}{DardhaG18}
\bibitem{DardhaG18}
\bibinfo{author}{Ornela \surnamestart Dardha\surnameend} \&
  \bibinfo{author}{Simon~J. \surnamestart Gay\surnameend}
  (\bibinfo{year}{2018}): \emph{\bibinfo{title}{A New Linear Logic for
  Deadlock-Free Session-Typed Processes}}.
\newblock In \bibinfo{editor}{Christel \surnamestart Baier\surnameend} \&
  \bibinfo{editor}{Ugo~Dal \surnamestart Lago\surnameend}, editors: {\slshape
  \bibinfo{booktitle}{{FOSSACS}}}, {\slshape \bibinfo{series}{LNCS}}
  \bibinfo{volume}{10803}, \bibinfo{publisher}{Springer}, pp.
  \bibinfo{pages}{91--109}, \doi{10.1007/978-3-319-89366-2\_5}.

\bibitemdeclare{inproceedings}{DGS12}
\bibitem{DGS12}
\bibinfo{author}{Ornela \surnamestart Dardha\surnameend},
  \bibinfo{author}{Elena \surnamestart Giachino\surnameend} \&
  \bibinfo{author}{Davide \surnamestart Sangiorgi\surnameend}
  (\bibinfo{year}{2012}): \emph{\bibinfo{title}{Session types revisited}}.
\newblock In: {\slshape \bibinfo{booktitle}{PPDP}}, \bibinfo{publisher}{ACM},
  \bibinfo{address}{New York, NY, USA}, pp. \bibinfo{pages}{139--150},
  \doi{10.1145/2370776.2370794}.

\bibitemdeclare{article}{DGS17}
\bibitem{DGS17}
\bibinfo{author}{Ornela \surnamestart Dardha\surnameend},
  \bibinfo{author}{Elena \surnamestart Giachino\surnameend} \&
  \bibinfo{author}{Davide \surnamestart Sangiorgi\surnameend}
  (\bibinfo{year}{2017}): \emph{\bibinfo{title}{Session types revisited}}.
\newblock {\slshape \bibinfo{journal}{Inf. Comput.}} \bibinfo{volume}{256}, pp.
  \bibinfo{pages}{253--286}, \doi{10.1016/j.ic.2017.06.002}.

\bibitemdeclare{article}{DP22}
\bibitem{DP22}
\bibinfo{author}{Ornela \surnamestart Dardha\surnameend} \&
  \bibinfo{author}{Jorge~A. \surnamestart P{\'{e}}rez\surnameend}
  (\bibinfo{year}{2022}): \emph{\bibinfo{title}{Comparing type systems for
  deadlock freedom}}.
\newblock {\slshape \bibinfo{journal}{J. Log. Algebraic Methods Program.}}
  \bibinfo{volume}{124}, p. \bibinfo{pages}{100717},
  \doi{10.1016/j.jlamp.2021.100717}.

\bibitemdeclare{inproceedings}{encoding}
\bibitem{encoding}
\bibinfo{author}{Romain \surnamestart Demangeon\surnameend} \&
  \bibinfo{author}{Kohei \surnamestart Honda\surnameend}
  (\bibinfo{year}{2011}): \emph{\bibinfo{title}{Full Abstraction in a Subtyped
  pi-Calculus with Linear Types}}.
\newblock In: {\slshape \bibinfo{booktitle}{{CONCUR}}}, {\slshape
  \bibinfo{series}{LNCS}} \bibinfo{volume}{6901},
  \bibinfo{publisher}{Springer}, pp. \bibinfo{pages}{280--296},
  \doi{10.1007/978-3-642-23217-6_19}.

\bibitemdeclare{inproceedings}{DGDY06}
\bibitem{DGDY06}
\bibinfo{author}{Mariangiola \surnamestart Dezani{-}Ciancaglini\surnameend},
  \bibinfo{author}{Elena \surnamestart Giachino\surnameend},
  \bibinfo{author}{Sophia \surnamestart Drossopoulou\surnameend} \&
  \bibinfo{author}{Nobuko \surnamestart Yoshida\surnameend}
  (\bibinfo{year}{2007}): \emph{\bibinfo{title}{Bounded Session Types for
  Object Oriented Languages}}.
\newblock In: {\slshape \bibinfo{booktitle}{FMCO}}, {\slshape
  \bibinfo{series}{LNCS}} \bibinfo{volume}{4709},
  \bibinfo{publisher}{Springer}, pp. \bibinfo{pages}{207--245},
  \doi{10.1007/978-3-540-74792-5_10}.

\bibitemdeclare{inproceedings}{DMYD06}
\bibitem{DMYD06}
\bibinfo{author}{Mariangiola \surnamestart Dezani-Ciancaglini\surnameend},
  \bibinfo{author}{Dimitris \surnamestart Mostrous\surnameend},
  \bibinfo{author}{Nobuko \surnamestart Yoshida\surnameend} \&
  \bibinfo{author}{Sophia \surnamestart Drossopoulou\surnameend}
  (\bibinfo{year}{2006}): \emph{\bibinfo{title}{Session Types for
  Object-Oriented Languages}}.
\newblock In: {\slshape \bibinfo{booktitle}{ECOOP 2006}}, {\slshape
  \bibinfo{series}{{LNCS}}} \bibinfo{volume}{4067},
  \bibinfo{publisher}{{Springer}}, pp. \bibinfo{pages}{328--352},
  \doi{10.1007/11785477_20}.

\bibitemdeclare{inproceedings}{KDLM21}
\bibitem{KDLM21}
\bibinfo{author}{Simon \surnamestart Fowler\surnameend}, \bibinfo{author}{Wen
  \surnamestart Kokke\surnameend}, \bibinfo{author}{Ornela \surnamestart
  Dardha\surnameend}, \bibinfo{author}{Sam \surnamestart Lindley\surnameend} \&
  \bibinfo{author}{J.~Garrett \surnamestart Morris\surnameend}
  (\bibinfo{year}{2021}): \emph{\bibinfo{title}{Separating Sessions Smoothly}}.
\newblock In \bibinfo{editor}{Serge \surnamestart Haddad\surnameend} \&
  \bibinfo{editor}{Daniele \surnamestart Varacca\surnameend}, editors:
  {\slshape \bibinfo{booktitle}{{CONCUR}}}, {\slshape \bibinfo{series}{LIPIcs}}
  \bibinfo{volume}{203}, \bibinfo{publisher}{Schloss Dagstuhl - Leibniz-Zentrum
  f{\"{u}}r Informatik}, pp. \bibinfo{pages}{36:1--36:18},
  \doi{10.4230/LIPIcs.CONCUR.2021.36}.

\bibitemdeclare{article}{boundpoly}
\bibitem{boundpoly}
\bibinfo{author}{Simon~J. \surnamestart Gay\surnameend} (\bibinfo{year}{2008}):
  \emph{\bibinfo{title}{Bounded polymorphism in session types}}.
\newblock {\slshape \bibinfo{journal}{Mathematical Structures in Computer
  Science}} \bibinfo{volume}{18}(\bibinfo{number}{5}), pp.
  \bibinfo{pages}{895--930}, \doi{10.1017/S0960129508006944}.

\bibitemdeclare{article}{gayhole}
\bibitem{gayhole}
\bibinfo{author}{Simon~J. \surnamestart Gay\surnameend} \&
  \bibinfo{author}{Malcolm \surnamestart Hole\surnameend}
  (\bibinfo{year}{2005}): \emph{\bibinfo{title}{Subtyping for session types in
  the pi calculus}}.
\newblock {\slshape \bibinfo{journal}{Acta Inf.}}
  \bibinfo{volume}{42}(\bibinfo{number}{2-3}), pp. \bibinfo{pages}{191--225},
  \doi{10.1007/s00236-005-0177-z}.

\bibitemdeclare{inproceedings}{GayThiemannVasconcelos20}
\bibitem{GayThiemannVasconcelos20}
\bibinfo{author}{Simon~J. \surnamestart Gay\surnameend}, \bibinfo{author}{Peter
  \surnamestart Thiemann\surnameend} \& \bibinfo{author}{Vasco~T. \surnamestart
  Vasconcelos\surnameend} (\bibinfo{year}{2020}): \emph{\bibinfo{title}{Duality
  of Session Types: The Final Cut}}.
\newblock In \bibinfo{editor}{Stephanie \surnamestart Balzer\surnameend} \&
  \bibinfo{editor}{Luca \surnamestart Padovani\surnameend}, editors: {\slshape
  \bibinfo{booktitle}{{PLACES@ETAPS}}}, {\slshape \bibinfo{series}{{EPTCS}}}
  \bibinfo{volume}{314}, pp. \bibinfo{pages}{23--33},
  \doi{10.4204/EPTCS.314.3}.

\bibitemdeclare{article}{GV10}
\bibitem{GV10}
\bibinfo{author}{Simon~J. \surnamestart Gay\surnameend} \&
  \bibinfo{author}{Vasco~Thudichum \surnamestart Vasconcelos\surnameend}
  (\bibinfo{year}{2010}): \emph{\bibinfo{title}{Linear type theory for
  asynchronous session types}}.
\newblock {\slshape \bibinfo{journal}{J. Funct. Program.}}
  \bibinfo{volume}{20}(\bibinfo{number}{1}), pp. \bibinfo{pages}{19--50},
  \doi{10.1017/S0956796809990268}.

\bibitemdeclare{article}{Hewitt77}
\bibitem{Hewitt77}
\bibinfo{author}{Carl \surnamestart Hewitt\surnameend} (\bibinfo{year}{1977}):
  \emph{\bibinfo{title}{Viewing Control Structures as Patterns of Passing
  Messages}}.
\newblock {\slshape \bibinfo{journal}{Artif. Intell.}}
  \bibinfo{volume}{8}(\bibinfo{number}{3}), pp. \bibinfo{pages}{323--364},
  \doi{10.1016/0004-3702(77)90033-9}.

\bibitemdeclare{inproceedings}{H93}
\bibitem{H93}
\bibinfo{author}{Kohei \surnamestart Honda\surnameend} (\bibinfo{year}{1993}):
  \emph{\bibinfo{title}{Types for Dyadic Interaction}}.
\newblock In: {\slshape \bibinfo{booktitle}{CONCUR}}, {\slshape
  \bibinfo{series}{LNCS}} \bibinfo{volume}{715}, \bibinfo{publisher}{Springer},
  pp. \bibinfo{pages}{509--523}, \doi{10.1007/3-540-57208-2_35}.

\bibitemdeclare{inproceedings}{kubo2}
\bibitem{kubo2}
\bibinfo{author}{Kohei \surnamestart Honda\surnameend}, \bibinfo{author}{Vasco
  \surnamestart Vasconcelos\surnameend} \& \bibinfo{author}{Makoto
  \surnamestart Kubo\surnameend} (\bibinfo{year}{1998}):
  \emph{\bibinfo{title}{Language primitives and type disciplines for structured
  communication-based programming}}.
\newblock In: {\slshape \bibinfo{booktitle}{ESOP}}, {\slshape
  \bibinfo{series}{{LNCS}}} \bibinfo{volume}{1381},
  \bibinfo{publisher}{{Springer}}, pp. \bibinfo{pages}{22--138},
  \doi{10.1007/BFb0053567}.

\bibitemdeclare{inproceedings}{HYC08}
\bibitem{HYC08}
\bibinfo{author}{Kohei \surnamestart Honda\surnameend}, \bibinfo{author}{Nobuko
  \surnamestart Yoshida\surnameend} \& \bibinfo{author}{Marco \surnamestart
  Carbone\surnameend} (\bibinfo{year}{2008}): \emph{\bibinfo{title}{Multiparty
  asynchronous session types}}.
\newblock In: {\slshape \bibinfo{booktitle}{POPL}}, \bibinfo{volume}{43(1)},
  \bibinfo{publisher}{ACM}, pp. \bibinfo{pages}{273--284},
  \doi{10.1145/1328438.1328472}.

\bibitemdeclare{article}{HYC16}
\bibitem{HYC16}
\bibinfo{author}{Kohei \surnamestart Honda\surnameend}, \bibinfo{author}{Nobuko
  \surnamestart Yoshida\surnameend} \& \bibinfo{author}{Marco \surnamestart
  Carbone\surnameend} (\bibinfo{year}{2016}): \emph{\bibinfo{title}{Multiparty
  asynchronous session types}}.
\newblock {\slshape \bibinfo{journal}{Journal of the ACM}}
  \bibinfo{volume}{63}(\bibinfo{number}{1}), p.~\bibinfo{pages}{9},
  \doi{10.1145/2827695}.

\bibitemdeclare{article}{K02}
\bibitem{K02}
\bibinfo{author}{Naoki \surnamestart Kobayashi\surnameend}
  (\bibinfo{year}{2002}): \emph{\bibinfo{title}{A Type System for Lock-Free
  Processes}}.
\newblock {\slshape \bibinfo{journal}{Inf. Comput.}}
  \bibinfo{volume}{177}(\bibinfo{number}{2}), pp. \bibinfo{pages}{122--159},
  \doi{10.1006/inco.2002.3171}.

\bibitemdeclare{inproceedings}{kobap}
\bibitem{kobap}
\bibinfo{author}{Naoki \surnamestart Kobayashi\surnameend}
  (\bibinfo{year}{2002}): \emph{\bibinfo{title}{Type Systems for Concurrent
  Programs}}.
\newblock In: {\slshape \bibinfo{booktitle}{10th Anniversary Colloquium of
  UNU/IIST}}, pp. \bibinfo{pages}{439--453},
  \doi{10.1007/978-3-540-40007-3_26}.

\bibitemdeclare{inproceedings}{deadlock2}
\bibitem{deadlock2}
\bibinfo{author}{Naoki \surnamestart Kobayashi\surnameend}
  (\bibinfo{year}{2006}): \emph{\bibinfo{title}{A New Type System for
  Deadlock-Free Processes}}.
\newblock In: {\slshape \bibinfo{booktitle}{CONCUR}}, {\slshape
  \bibinfo{series}{LNCS}} \bibinfo{volume}{4137},
  \bibinfo{publisher}{Springer}, pp. \bibinfo{pages}{233--247},
  \doi{10.1007/11817949_16}.

\bibitemdeclare{unpublished}{koba}
\bibitem{koba}
\bibinfo{author}{Naoki \surnamestart Kobayashi\surnameend}
  (\bibinfo{year}{2007}): \emph{\bibinfo{title}{Type Systems for Concurrent
  Programs}}.
\newblock
  \urlprefix\url{http://www.kb.ecei.tohoku.ac.jp/~koba/papers/tutorial-type-extended.pdf}.
\newblock \bibinfo{note}{Extended version of \cite{kobap}, Tohoku University}.

\bibitemdeclare{article}{linearity}
\bibitem{linearity}
\bibinfo{author}{Naoki \surnamestart Kobayashi\surnameend},
  \bibinfo{author}{Benjamin~C. \surnamestart Pierce\surnameend} \&
  \bibinfo{author}{David~N. \surnamestart Turner\surnameend}
  (\bibinfo{year}{1999}): \emph{\bibinfo{title}{Linearity and the
  pi-calculus}}.
\newblock {\slshape \bibinfo{journal}{ACM Trans. Program. Lang. Syst.}}
  \bibinfo{volume}{21}(\bibinfo{number}{5}), pp. \bibinfo{pages}{914--947},
  \doi{10.1145/330249.330251}.

\bibitemdeclare{inproceedings}{KD21}
\bibitem{KD21}
\bibinfo{author}{Wen \surnamestart Kokke\surnameend} \& \bibinfo{author}{Ornela
  \surnamestart Dardha\surnameend} (\bibinfo{year}{2021}):
  \emph{\bibinfo{title}{Deadlock-free session types in linear Haskell}}.
\newblock In \bibinfo{editor}{Jurriaan \surnamestart Hage\surnameend}, editor:
  {\slshape \bibinfo{booktitle}{Haskell}}, \bibinfo{publisher}{{ACM}}, pp.
  \bibinfo{pages}{1--13}, \doi{10.1145/3471874.3472979}.

\bibitemdeclare{inproceedings}{KokkeD21}
\bibitem{KokkeD21}
\bibinfo{author}{Wen \surnamestart Kokke\surnameend} \& \bibinfo{author}{Ornela
  \surnamestart Dardha\surnameend} (\bibinfo{year}{2021}):
  \emph{\bibinfo{title}{Prioritise the Best Variation}}.
\newblock In \bibinfo{editor}{Kirstin \surnamestart Peters\surnameend} \&
  \bibinfo{editor}{Tim A.~C. \surnamestart Willemse\surnameend}, editors:
  {\slshape \bibinfo{booktitle}{{FORTE}}}, {\slshape \bibinfo{series}{LNCS}}
  \bibinfo{volume}{12719}, \bibinfo{publisher}{Springer}, pp.
  \bibinfo{pages}{100--119}, \doi{10.1007/978-3-030-78089-0\_6}.

\bibitemdeclare{inproceedings}{lindleymorris16}
\bibitem{lindleymorris16}
\bibinfo{author}{Sam \surnamestart Lindley\surnameend} \&
  \bibinfo{author}{J.~Garrett \surnamestart Morris\surnameend}
  (\bibinfo{year}{2016}): \emph{\bibinfo{title}{Embedding session types in
  Haskell}}.
\newblock In: {\slshape \bibinfo{booktitle}{Proc.\ of Haskell}},
  \bibinfo{publisher}{{ACM}}, pp. \bibinfo{pages}{133--145},
  \doi{10.1145/2976002.2976018}.

\bibitemdeclare{inproceedings}{MY13}
\bibitem{MY13}
\bibinfo{author}{Fabrizio \surnamestart Montesi\surnameend} \&
  \bibinfo{author}{Nobuko \surnamestart Yoshida\surnameend}
  (\bibinfo{year}{2013}): \emph{\bibinfo{title}{Compositional Choreographies}}.
\newblock In: {\slshape \bibinfo{booktitle}{CONCUR}}, {\slshape
  \bibinfo{series}{LNCS}} \bibinfo{volume}{8052},
  \bibinfo{publisher}{Springer}, pp. \bibinfo{pages}{425--439},
  \doi{10.1007/978-3-642-40184-8_30}.

\bibitemdeclare{inproceedings}{ho}
\bibitem{ho}
\bibinfo{author}{Dimitris \surnamestart Mostrous\surnameend} \&
  \bibinfo{author}{Nobuko \surnamestart Yoshida\surnameend}
  (\bibinfo{year}{2007}): \emph{\bibinfo{title}{Two Session Typing Systems for
  Higher-Order Mobile Processes}}.
\newblock In: {\slshape \bibinfo{booktitle}{TLCA}}, {\slshape
  \bibinfo{series}{LNCS}} \bibinfo{volume}{4583},
  \bibinfo{publisher}{Springer}, pp. \bibinfo{pages}{321--335},
  \doi{10.1007/978-3-540-73228-0_23}.

\bibitemdeclare{article}{orchardyoshida17}
\bibitem{orchardyoshida17}
\bibinfo{author}{Dominic \surnamestart Orchard\surnameend} \&
  \bibinfo{author}{Nobuko \surnamestart Yoshida\surnameend}
  (\bibinfo{year}{2017}): \emph{\bibinfo{title}{{Session Types with Linearity
  in Haskell}}}.
\newblock {\slshape \bibinfo{journal}{Behavioural Types: from Theory to
  Tools}}, pp. \bibinfo{pages}{219--242}, \doi{10.13052/rp-9788793519817}.

\bibitemdeclare{inproceedings}{Padovani14}
\bibitem{Padovani14}
\bibinfo{author}{Luca \surnamestart Padovani\surnameend}
  (\bibinfo{year}{2014}): \emph{\bibinfo{title}{Deadlock and lock freedom in
  the linear {\(\pi\)}-calculus}}.
\newblock In \bibinfo{editor}{Thomas~A. \surnamestart Henzinger\surnameend} \&
  \bibinfo{editor}{Dale \surnamestart Miller\surnameend}, editors: {\slshape
  \bibinfo{booktitle}{{CSL-LICS}}}, \bibinfo{publisher}{{ACM}}, pp.
  \bibinfo{pages}{72:1--72:10}, \doi{10.1145/2603088.2603116}.

\bibitemdeclare{incollection}{Padovani17}
\bibitem{Padovani17}
\bibinfo{author}{Luca \surnamestart Padovani\surnameend}
  (\bibinfo{year}{2017}): \emph{\bibinfo{title}{Type-Based Analysis of Linear
  Communications}}.
\newblock In \bibinfo{editor}{Simon \surnamestart Gay\surnameend} \&
  \bibinfo{editor}{Ant\'onio \surnamestart Ravara\surnameend}, editors:
  {\slshape \bibinfo{booktitle}{Behavioural Types: from Theory to Tools}},
  \bibinfo{publisher}{River Publishers}, pp. \bibinfo{pages}{193--217},
  \doi{10.13052/rp-9788793519817}.

\bibitemdeclare{inproceedings}{sub}
\bibitem{sub}
\bibinfo{author}{Benjamin~C. \surnamestart Pierce\surnameend} \&
  \bibinfo{author}{Davide \surnamestart Sangiorgi\surnameend}
  (\bibinfo{year}{1993}): \emph{\bibinfo{title}{Typing and Subtyping for Mobile
  Processes}}.
\newblock In: {\slshape \bibinfo{booktitle}{LICS}}, \bibinfo{publisher}{{IEEE}
  Computer Society}, pp. \bibinfo{pages}{376--385},
  \doi{10.1109/LICS.1993.287570}.

\bibitemdeclare{inproceedings}{pucellatov08}
\bibitem{pucellatov08}
\bibinfo{author}{Riccardo \surnamestart Pucella\surnameend} \&
  \bibinfo{author}{Jesse~A. \surnamestart Tov\surnameend}
  (\bibinfo{year}{2008}): \emph{\bibinfo{title}{Haskell session types with
  (almost) no class}}.
\newblock In: {\slshape \bibinfo{booktitle}{Proc.\ of Haskell}},
  \bibinfo{publisher}{{ACM}}, \doi{10.1145/1411286.1411290}.

\bibitemdeclare{article}{S98}
\bibitem{S98}
\bibinfo{author}{Davide \surnamestart Sangiorgi\surnameend}
  (\bibinfo{year}{1998}): \emph{\bibinfo{title}{An Interpretation of Typed
  Objects into Typed pi-Calculus}}.
\newblock {\slshape \bibinfo{journal}{Inf. Comput.}}
  \bibinfo{volume}{143}(\bibinfo{number}{1}), pp. \bibinfo{pages}{34--73},
  \doi{10.1006/inco.1998.2711}.

\bibitemdeclare{book}{pi}
\bibitem{pi}
\bibinfo{author}{Davide \surnamestart Sangiorgi\surnameend} \&
  \bibinfo{author}{David \surnamestart Walker\surnameend}
  (\bibinfo{year}{2001}): \emph{\bibinfo{title}{The Pi-Calculus - a theory of
  mobile processes}}.
\newblock \bibinfo{publisher}{Cambridge University Press}.

\bibitemdeclare{inproceedings}{SDHY17}
\bibitem{SDHY17}
\bibinfo{author}{Alceste \surnamestart Scalas\surnameend},
  \bibinfo{author}{Ornela \surnamestart Dardha\surnameend},
  \bibinfo{author}{Raymond \surnamestart Hu\surnameend} \&
  \bibinfo{author}{Nobuko \surnamestart Yoshida\surnameend}
  (\bibinfo{year}{2017}): \emph{\bibinfo{title}{A Linear Decomposition of
  Multiparty Sessions for Safe Distributed Programming}}.
\newblock In \bibinfo{editor}{Peter \surnamestart M{\"{u}}ller\surnameend},
  editor: {\slshape \bibinfo{booktitle}{{ECOOP}}}, {\slshape
  \bibinfo{series}{LIPIcs}}~\bibinfo{volume}{74}, \bibinfo{publisher}{Schloss
  Dagstuhl - Leibniz-Zentrum f{\"{u}}r Informatik}, pp.
  \bibinfo{pages}{24:1--24:31}, \doi{10.4230/LIPIcs.ECOOP.2017.24}.

\bibitemdeclare{techreport}{TECHREPORT}
\bibitem{TECHREPORT}
\bibinfo{author}{Alceste \surnamestart Scalas\surnameend},
  \bibinfo{author}{Ornela \surnamestart Dardha\surnameend},
  \bibinfo{author}{Raymond \surnamestart Hu\surnameend} \&
  \bibinfo{author}{Nobuko \surnamestart Yoshida\surnameend}
  (\bibinfo{year}{2017}): \emph{\bibinfo{title}{A Linear Decomposition of
  Multiparty Sessions for Safe Distributed Programming}}.
\newblock \bibinfo{type}{Technical Report} \bibinfo{number}{2},
  \bibinfo{institution}{{Imperial} {College} {London}}.
\newblock
  \urlprefix\url{https://www.doc.ic.ac.uk/research/technicalreports/2017/#2}.

\bibitemdeclare{article}{SDHY17b}
\bibitem{SDHY17b}
\bibinfo{author}{Alceste \surnamestart Scalas\surnameend},
  \bibinfo{author}{Ornela \surnamestart Dardha\surnameend},
  \bibinfo{author}{Raymond \surnamestart Hu\surnameend} \&
  \bibinfo{author}{Nobuko \surnamestart Yoshida\surnameend}
  (\bibinfo{year}{2017}): \emph{\bibinfo{title}{A Linear Decomposition of
  Multiparty Sessions for Safe Distributed Programming (Artifact)}}.
\newblock {\slshape \bibinfo{journal}{Dagstuhl Artifacts Ser.}}
  \bibinfo{volume}{3}(\bibinfo{number}{2}), pp. \bibinfo{pages}{03:1--03:2},
  \doi{10.4230/DARTS.3.2.3}.

\bibitemdeclare{inproceedings}{ScalasY16}
\bibitem{ScalasY16}
\bibinfo{author}{Alceste \surnamestart Scalas\surnameend} \&
  \bibinfo{author}{Nobuko \surnamestart Yoshida\surnameend}
  (\bibinfo{year}{2016}): \emph{\bibinfo{title}{Lightweight Session Programming
  in Scala}}.
\newblock In \bibinfo{editor}{Shriram \surnamestart Krishnamurthi\surnameend}
  \& \bibinfo{editor}{Benjamin~S. \surnamestart Lerner\surnameend}, editors:
  {\slshape \bibinfo{booktitle}{{ECOOP}}}, {\slshape
  \bibinfo{series}{LIPIcs}}~\bibinfo{volume}{56}, \bibinfo{publisher}{Schloss
  Dagstuhl - Leibniz-Zentrum f{\"{u}}r Informatik}, pp.
  \bibinfo{pages}{21:1--21:28}, \doi{10.4230/LIPIcs.ECOOP.2016.21}.

\bibitemdeclare{inproceedings}{kubo1}
\bibitem{kubo1}
\bibinfo{author}{Kaku \surnamestart Takeuchi\surnameend},
  \bibinfo{author}{Kohei \surnamestart Honda\surnameend} \&
  \bibinfo{author}{Makoto \surnamestart Kubo\surnameend}
  (\bibinfo{year}{1994}): \emph{\bibinfo{title}{An Interaction-based Language
  and its Typing System}}.
\newblock In: {\slshape \bibinfo{booktitle}{PARLE}}, {\slshape
  \bibinfo{series}{LNCS}} \bibinfo{volume}{817}, \bibinfo{publisher}{Springer},
  pp. \bibinfo{pages}{398--413}, \doi{10.1007/3-540-58184-7_118}.

\bibitemdeclare{article}{VVR06}
\bibitem{VVR06}
\bibinfo{author}{Antonio \surnamestart Vallecillo\surnameend},
  \bibinfo{author}{Vasco~Thudichum \surnamestart Vasconcelos\surnameend} \&
  \bibinfo{author}{Ant{\'o}nio \surnamestart Ravara\surnameend}
  (\bibinfo{year}{2006}): \emph{\bibinfo{title}{Typing the Behavior of Software
  Components using Session Types}}.
\newblock {\slshape \bibinfo{journal}{Fundam. Inform.}}
  \bibinfo{volume}{73}(\bibinfo{number}{4}), pp. \bibinfo{pages}{583--598}.
\newblock
  \urlprefix\url{https://content.iospress.com/articles/fundamenta-informaticae/fi73-4-07}.

\bibitemdeclare{article}{fundamentals}
\bibitem{fundamentals}
\bibinfo{author}{Vasco~T. \surnamestart Vasconcelos\surnameend}
  (\bibinfo{year}{2012}): \emph{\bibinfo{title}{Fundamentals of session
  types}}.
\newblock {\slshape \bibinfo{journal}{Information Computation}}
  \bibinfo{volume}{217}, pp. \bibinfo{pages}{52--70},
  \doi{10.1016/j.ic.2012.05.002}.

\bibitemdeclare{article}{VGR06}
\bibitem{VGR06}
\bibinfo{author}{Vasco~Thudichum \surnamestart Vasconcelos\surnameend},
  \bibinfo{author}{Simon~J. \surnamestart Gay\surnameend} \&
  \bibinfo{author}{Ant{\'o}nio \surnamestart Ravara\surnameend}
  (\bibinfo{year}{2006}): \emph{\bibinfo{title}{Type checking a multithreaded
  functional language with session types}}.
\newblock {\slshape \bibinfo{journal}{Theor. Comput. Sci.}}
  \bibinfo{volume}{368}(\bibinfo{number}{1-2}), pp. \bibinfo{pages}{64--87},
  \doi{10.1016/j.tcs.2006.06.028}.

\bibitemdeclare{inproceedings}{W12}
\bibitem{W12}
\bibinfo{author}{Philip \surnamestart Wadler\surnameend}
  (\bibinfo{year}{2012}): \emph{\bibinfo{title}{Propositions as sessions}}.
\newblock In: {\slshape \bibinfo{booktitle}{ICFP}}, \bibinfo{publisher}{{ACM}},
  pp. \bibinfo{pages}{273--286}, \doi{10.1145/2364527.2364568}.

\bibitemdeclare{article}{revisited}
\bibitem{revisited}
\bibinfo{author}{Nobuko \surnamestart Yoshida\surnameend} \&
  \bibinfo{author}{Vasco~Thudichum \surnamestart Vasconcelos\surnameend}
  (\bibinfo{year}{2007}): \emph{\bibinfo{title}{Language Primitives and Type
  Discipline for Structured Communication-Based Programming Revisited: Two
  Systems for Higher-Order Session Communication}}.
\newblock {\slshape \bibinfo{journal}{Electr. Notes Theor. Comput. Sci.}}
  \bibinfo{volume}{171}(\bibinfo{number}{4}), pp. \bibinfo{pages}{73--93},
  \doi{10.1016/j.entcs.2007.02.056}.

\end{thebibliography}

\end{document}